\documentclass[8pt]{article}
\usepackage{a4wide,epsfig}

\usepackage{graphicx}
\usepackage{amsmath}
\usepackage{slashed}
\usepackage{graphicx} 
\usepackage{caption,subcaption}
\usepackage{axodraw4j}
\usepackage{color}

\newcommand{\agt}{\rlap{\lower 3.5 pt \hbox{$\mathchar \sim$}} \raise 1pt
 \hbox {$>$}}
\newcommand{\alt}{\rlap{\lower 3.5 pt \hbox{$\mathchar \sim$}} \raise 1pt
 \hbox {$<$}}

\catcode`@=11
\newcount\@tempcntc
\def\@citex[#1]#2{\if@filesw\immediate\write\@auxout{\string\citation{#2}}\fi
  \@tempcnta\z@\@tempcntb\m@ne\def\@citea{}\@cite{\@for\@citeb:=#2\do
    {\@ifundefined
       {b@\@citeb}{\@citeo\@tempcntb\m@ne\@citea\def\@citea{,}{\bf
?}\@warning
       {Citation `\@citeb' on page \thepage \space undefined}}%
    {\setbox\z@\hbox{\global\@tempcntc0\csname b@\@citeb\endcsname\relax}%
     \ifnum\@tempcntc=\z@ \@citeo\@tempcntb\m@ne
       \@citea\def\@citea{,}\hbox{\csname b@\@citeb\endcsname}%
     \else
      \advance\@tempcntb\@ne
      \ifnum\@tempcntb=\@tempcntc
      \else\advance\@tempcntb\m@ne\@citeo
      \@tempcnta\@tempcntc\@tempcntb\@tempcntc\fi\fi}}\@citeo}{#1}}
\def\@citeo{\ifnum\@tempcnta>\@tempcntb\else\@citea\def\@citea{,}%
  \ifnum\@tempcnta=\@tempcntb\the\@tempcnta\else
   {\advance\@tempcnta\@ne\ifnum\@tempcnta=\@tempcntb \else
\def\@citea{--}\fi
    \advance\@tempcnta\m@ne\the\@tempcnta\@citea\the\@tempcntb}\fi\fi}
\catcode`@=12

\begin{document}

\title{
\vskip-3cm{\baselineskip14pt
\centerline{\normalsize DESY 20--212\hfill ISSN 0418-9833}
\centerline{\normalsize December 2022\hfill}}
\vskip1.5cm
Matching the Standard Model to Heavy-Quark Effective Theory and Nonrelativistic QCD}

\author{Beno\^{i}t~Assi,\footnote{Present address: Theory Division, Fermilab,
PO Box 500, MS 106, Batavia, IL~60510-5011, USA.}\ \
Bernd A. Kniehl\\
{\normalsize II. Institut f\"ur Theoretische Physik, Universit\"at Hamburg,}\\
{\normalsize Luruper Chaussee 149, 22761 Hamburg, Germany}\\
Joan~Soto\\
  {\normalsize Departament de F\'\i sica Qu\`antica i Astrof\'isica and Institut de Ci\`encies del Cosmos,}\\
{\normalsize  Universitat de Barcelona, Mart\'\i$\,$ i Franqu\`es 1, 08028 Barcelona, Catalonia, Spain}
}

\date{\today}

\maketitle

\begin{abstract}
We find the leading electroweak corrections to the Lagrangians of heavy-quark effective theory and nonrelativistic QCD. These corrections appear in the Wilson coefficients of the two- and four-quark operators and are considered here at one-loop order through $\mathcal{O}(1/m^3)$ and $\mathcal{O}(1/m^2)$, respectively. The two-quark operators through this order include new parity violating terms, which we derive analogously to the parity preserving QCD result at one-loop order.
%
\end{abstract}

\section{Introduction}
\label{sec:intro}
Effective field theories (EFTs) are important tools in elementary particle
physics and, in particular, in quantum chromodynamics (QCD).
They allow for an economic, yet precise treatment of problems that involve
widely separated mass scales.
In many important applications of QCD, the mass of a heavy quark (top, bottom,
or charm) is much larger than the remaining dynamical scales of the problem
considered.
Heavy-quark effective theory (HQET) and nonrelativistic QCD (NRQCD), which are
among the most frequently used EFTs of QCD, are tailored for such systems.
It is the purpose of this paper to extend HQET and NRQCD from pure QCD to the
full Standard Model (SM).

More specifically, HQET has mainly been employed to study systems involving one
heavy quark $Q$~\cite{Shifman:1987rj,Politzer:1988wp,Isgur:1989vq}.
In these studies, when considering heavy-light systems, the authors reduce the
problem down to one with two dynamical scales: the heavy-quark mass, $m$, and
the scale of the rest, which is chosen to be the quark confinement scale,
$\Lambda_\mathrm{QCD}$.
One then constructs the HQET Lagrangian as a power series in the inverse
heavy-quark pole mass.
One can then estimate the size of each term by assigning the scale
$\Lambda_\mathrm{QCD}$ to every parameter present other than the heavy-quark mass.
One is thus left with operators exhibiting two distinct structures: terms
containing light degrees of freedom describing gluons and light quarks and
terms that are bi-linear in the heavy-quark fields.

On the other hand, NRQCD is mostly employed to study systems involving a
heavy quark-antiquark bound state, $Q\Bar{Q}$~\cite{Caswell:1985ui,Bodwin:1994jh}.
In NRQCD, one usually takes into account two additional dynamical scales: the
relative momentum, $|\boldsymbol{q}|\sim mv$, where $v$ is
the relative velocity of $Q$ and $\Bar{Q}$ in the $Q\Bar{Q}$ rest frame, and the
binding energy, $E\sim mv^2$, of the $Q\Bar{Q}$ bound state.
These extra scales add increased complexity to the power counting rules.
Thus, the size of each term in the NRQCD Lagrangian is no longer unique, but
depends on the system under consideration.
One can, however, still provide reasonable estimates of the leading size of
each term by means of velocity scaling rules~\cite{Brambilla:1999xf,Bodwin:1994jh,Lepage:1992tx,Manohar:2000hj}.

The difference between HQET and NRQCD is immediately clear by considering the
first two bi-linear terms in the effective Lagrangian,
\begin{equation}
    \mathcal{L}=\psi^\dagger \left(iD^0+\frac{\boldsymbol{D}^2}{2m}\right)\psi\,.
\end{equation}
To compare the two theories, one can observe that, in HQET, the first and second terms are of orders $\mathcal{O}(\Lambda_\mathrm{QCD})$ and $\mathcal{O}(\Lambda_\mathrm{QCD}^2/m)$, respectively, while, in NRQCD, they are both of order
$\mathcal{O}(mv^2)$.
Thus, one can immediately understand that the heavy-quark propagator in HQET is $i/(k^0+i\epsilon)$ and in NRQCD it is $i/(k^0-\boldsymbol{k}^2/2m+i\epsilon)$. The NRQCD Lagrangian mimics the HQET Lagrangian in that it consists of terms in
a power series expansion in the heavy-quark mass.
It contains two- and four-quark operators, {\it i.e.}\ terms bi-linear in the
heavy-(anti)quark fields and terms bi-linear in both heavy-quark and heavy-antiquark fields, respectively. 

Our work is focused on calculating the primary building block of an EFT, the
EFT Lagrangian, and its matching to the full-theory Lagrangian.
The matching process is achievable by making sure that the full-theory and EFT
$S$ matrix elements are equal.
Both the NRQCD and HQET matching conditions are computed in the same way, and
the Lagrangians are thus identical~\cite{Manohar:1997qy}.
The parameters that are modified by the matching procedure are called the
matching (or Wilson) coefficients, which multiply the respective operators in
the EFT.
The matching in NRQCD is then achieved order by order in the strong-coupling
constant, $\alpha_s$, and the inverse heavy-quark mass~\cite{Pineda:1998kj}. 

In this paper, we will extend the NRQCD Lagrangian by taking into account the
leading electroweak (EW) corrections to the two- and four-quark operators of
NRQCD at one-loop order, retaining terms through orders
$\mathcal{O}(\alpha/m^3)$ and $\mathcal{O}(\alpha^2/m^2)$ respectively.
Although the Wilson
coefficients are known in the EFT through $\mathcal{O}(\alpha_s^2/m^2)$ \cite{Gerlach:2019kfo} and $\mathcal{O}(\alpha_s/m^5)$ \cite{Huang:2020fiu}, the EW corrections have not yet been considered in full detail.
They must be incorporated, since, at leading order, they start altering the
matching coefficients by amounts comparable to the higher-order QCD terms.
Therefore, we study the effect of incorporating the EW contributions at leading order and notice how the matching coefficients are improved.
Moreover, the Lagrangian itself must be extended to include parity violating
operators to enable the matching to the SM, as parity symmetry holds for QCD, but not for the full SM.
The usefulness of our efforts lies in the prolific use of heavy-quark EFTs for
high-precision predictions of observables at threshold energies, which would be
the primary purpose of a future $e^+e^-$ collider~\cite{Frey:1995ai}.
In particular, this includes the top-quark mass determination, which is crucial
for understanding the stability of the EW vacuum~\cite{Bezrukov:2012sa,Alekhin:2012py,Bednyakov:2015sca}.
Many so-called threshold quark mass definitions~\cite{Beneke:1998rk,Pineda:2001zq,Marquard:2015qpa} have arisen from the heavy-quark EFT
frameworks, and we know that the EW sector plays a crucial role in determining
the $\overline{\mathrm{MS}}$ mass of the top quark~\cite{Hempfling:1994ar,Kniehl:1994dz,Jegerlehner:2012kn}.
Thus, it stands to reason that the same is true for the threshold mass
definitions. 

This paper is organized as follows.
In Section~\ref{sec:Lagrangian}, we introduce our notation and write down the
effective Lagrangian.
In Section~\ref{sec:twoq}, we consider the bi-linear operators, evaluate the
various form factors, and perform the matching to find the Wilson coefficients.
In Section~\ref{sec:fourq}, we study the four-quark operators and extract
their Wilson coefficients for the cases of unequal and equal quark masses.
In Section~\ref{sec:discussion}, we undertake a detailed numerical analysis of
the EW radiative corrections and compare them with the pure QCD ones.
In Section~\ref{sec:conclusion}, we present our conclusions. 

\section{Lagrangian}
\label{sec:Lagrangian}

The continuum NRQCD Lagrangian, up to the order of interest here, has
previously been computed~\cite{Manohar:1997qy,Pineda:1998kj} using
dimensional regularization for the infrared (IR) and ultraviolet (UV)
divergences and taking the external states to be on mass shell.
To construct the NRQCD Lagrangian, one must consider heavy quarks and
antiquarks, with mass $m\gg\Lambda_\mathrm{QCD}$, coupled to non-Abelian gauge
fields, enforcing Hermicity, parity, time-reversal, and rotational invariance.
One can further perform heavy-quark field redefinitions to eliminate time
derivatives acting on the heavy-quark fields at higher orders in $1/m$.
This is known as the canonical form of the heavy-quark Lagrangian~\cite{Politzer:1980me}.
Notice that, when employing the NRQCD Lagrangian, which we define below, NRQCD
has UV cut-offs, $\nu_p$ and $\nu_s$, satisfying $mv\ll\nu_p,\nu_s\ll m$,
which corresponds to integrating out the hard modes of QCD to obtain NRQCD
\cite{Beneke:1997zp}.
Specifically, $\nu_p$ is the UV cut-off for the relative three-momentum
exchanged between the heavy quark and antiquark, and $\nu_s$ is the UV cut-off
for the three-momenta of the gluons and light quarks.
Up to field redefinitions, the NRQCD Lagrangian including light quarks reads
\cite{Caswell:1985ui,Eichten:1990vp,Brambilla:1999xf}:
\begin{equation}
  \mathcal{L}=\mathcal{L}_{\psi}+\mathcal{L}_{\chi}+\mathcal{L}_{\psi\chi}+\mathcal{L}_g+\mathcal{L}_l\,,
  \label{eq:nrqcdlagrangian}
\end{equation}
where $\psi$ and $\chi$ are the Pauli spinor fields that annihilate a heavy quark and create a heavy antiquark, respectively.
Specifically, $\mathcal{L}_{\psi}$ and $\mathcal{L}_{\chi}$ include the terms
bi-linear in the heavy-quark fields, $\mathcal{L}_{\psi\chi}$ accommodates the
four-quark operators, and $\mathcal{L}_g$ and $\mathcal{L}_l$ represent the
Yang-Mills and light-quark parts of the QCD Lagrangian, respectively. 
We are mainly interested in the first three terms on the right-hand side of
Eq.~(\ref{eq:nrqcdlagrangian}), as they attain the leading EW corrections to their matching coefficients.

More explicitly, working in a reference frame where $v^\mu=(1,\boldsymbol{0})$,
up to the order of interest here, we have~\cite{Caswell:1985ui,Manohar:1997qy,Kinoshita:1995mt},
\begin{eqnarray}
  \mathcal{L}_{\psi,\chi}&=&\psi^{\dagger}\left\lbrace i c_0 D_t+c_2\frac{\boldsymbol{D}^2}{2m}+c_4\frac{\boldsymbol{D}^4}{8m^3}+c_Fg_s\frac{\boldsymbol{\sigma}\cdot\boldsymbol{B}}{2m}
+  c_D g_s\frac{[\boldsymbol{D}\cdot\boldsymbol{E}]}{8m^2}
  + ic_S g_s\frac{\boldsymbol{\sigma}\cdot(\boldsymbol{D}\times\boldsymbol{E}-\boldsymbol{E}\times\boldsymbol{D})}{8m^2} \right.\nonumber\\ 
  &&{}+\left. c_{W_1}g_s\frac{\lbrace\boldsymbol{D^2},\boldsymbol{\sigma}\cdot\boldsymbol{B}\rbrace}{8m^3}-2c_{W_2}g_s\frac{\boldsymbol{D}_i\boldsymbol{\sigma}\cdot\boldsymbol{B}\boldsymbol{D}_i}{8m^3}
+c_q g_s\frac{\boldsymbol{\sigma}\cdot\boldsymbol{D}\boldsymbol{B}\cdot\boldsymbol{D}+\boldsymbol{D}\cdot\boldsymbol{B}\boldsymbol{\sigma}\cdot\boldsymbol{D}}{8m^3}  \right.\nonumber\\ 
  &&{}+\left. ic_Mg_s\frac{\boldsymbol{D}\cdot[\boldsymbol{D}\times\boldsymbol{B}]+[\boldsymbol{D}\times\boldsymbol{B}]\cdot\boldsymbol{D}}{8m^3} \right \rbrace\psi
+ (\mathrm{c.c.}, \psi\leftrightarrow\chi)  +\mathcal{O}\left(\frac{1}{m^4},\frac{g_s^2}{m^3}\right)\,,  
    \label{eqn:L2pt}
\end{eqnarray}
where $g_s$ is the QCD gauge coupling and c.c.\ stands for charge conjugate.
The terms in Eq.~(\ref{eqn:L2pt}) require some unpacking.
We decompose the covariant derivative,
$D_{\mu}=\partial_{\mu}+ig_sA_\mu^aT^a\equiv(D_t,\boldsymbol{D})$, with gluon
field $A_\mu^a$ and Gell-Mann matrices $T^a$, into time and spacial components,
$iD_t=i\partial_t -g_sA_0$ and
$i\boldsymbol{D}=i\boldsymbol{\partial}+g_s\boldsymbol{A}$.
Then, the combinations $\boldsymbol{E}=-\frac{i}{g_s}[D_t,\boldsymbol{D}]$ and
$B^i=\frac{i}{2g_s}\epsilon_{ijk}[D_j,D_k]$ are the QCD analogues of the electric and magnetic fields, respectively.
The subscripts $F$, $S$, and $D$ on the Wilson coefficients stand for Fermi,
spin-orbit, and Darwin, respectively.
We use the common summation convention, $X^iY^i\equiv \sum_{i=1}^3X^iY^i$, and
define $[X,Y]\equiv XY-YX$ and $\lbrace X,Y\rbrace\equiv XY+YX$ to denote
commutators and anticommutators, respectively.
Equations~(\ref{eqn:L2pt}) and (\ref{eqn:L4pt}) represent the most general
expressions that can be constructed from all possible rotationally invariant,
Hermitian combinations of $iD_t$, $\boldsymbol{D}$, $\boldsymbol{E}$, $i\boldsymbol{B}$, and $i\boldsymbol{\sigma}$, with parity requiring even numbers of factors of $\boldsymbol{D}$ and $\boldsymbol{E}$.

Furthermore, we have~\cite{Brambilla:2008zg}
\begin{eqnarray}
  \mathcal{L}_{\psi\chi}&=&\frac{d_{ss}}{m_1 m_2}\psi_1^{\dagger}\psi_1\chi_2^{\dagger}\chi_2+\frac{d_{sv}}{m_1 m_2}\psi_1^{\dagger}\boldsymbol{\sigma}\psi_1\chi_2^{\dagger}\boldsymbol{\sigma}\chi_2
  +\frac{d_{vs}}{m_1 m_2}\psi_1^{\dagger}T^a\psi_1\chi_2^{\dagger}T^a\chi_2
  +\frac{d_{vv}}{m_1m_2}\psi_1^{\dagger}T^a\boldsymbol{\sigma}\psi_1\chi_2^{\dagger}T^a\boldsymbol{\sigma}\chi_2\,, 
    \label{eqn:L4pt}
\end{eqnarray}
where we have allowed for different quark flavors, with masses $m_1$ and $m_2$.
The matching coefficients, $d_{xy}$, contain subindices which label the quark-antiquark states. Specifically, the first index corresponds to color ($s$ for singlet and $v$ for octet) and the second index refers to spin ($s$ for singlet and $v$ for triplet).

In fact, one can always rewrite the terms in Eq.~(\ref{eqn:L4pt}) via
identified Fiertz transformations~\cite{Dreiner:2008tw}.
In this way, $\mathcal{L}_{\psi\chi}$ can be cast into the alternative form
\begin{eqnarray}
  \mathcal{L}_{\psi\chi}&=&\frac{d^c_{ss}}{m_1 m_2}\psi_1^{\dagger}\chi_2\chi_2^{\dagger}\psi_1+\frac{d^c_{sv}}{m_1 m_2}\psi_1^{\dagger}\boldsymbol{\sigma}\chi_2\chi_2^{\dagger}\boldsymbol{\sigma}\psi_1
  +\frac{d^c_{vs}}{m_1 m_2}\psi_1^{\dagger}T^a\chi_2\chi_2^{\dagger}T^a\psi_1
  +\frac{d^c_{vs}}{m_1 m_2}\psi_1^{\dagger}T^a\boldsymbol{\sigma}\chi_2\chi_2^{\dagger}T^a\boldsymbol{\sigma}\psi_1\,, \label{eqn:L4pt2}
\end{eqnarray}
where the new basis of coefficient functions emerges from the old one via the
transformation
\begin{eqnarray}
    d_{ss}&=&-\frac{d_{ss}^c}{2N_c}-\frac{3d_{sv}^c}{2N_c}-\frac{N_c^2-1}{4N_c^2}d_{vs}^c-3\frac{N_c^2-1}{4N_c^2}d_{vv}^c,  \nonumber \\ 
    d_{sv}&=&-\frac{d_{ss}^c}{2N_c}+\frac{d_{sv}^c}{2N_c}-\frac{N_c^2-1}{4N_c^2}d_{vs}^c+\frac{N_c^2-1}{4N_c^2}d_{vv}^c,  \nonumber \\ 
    d_{vs}&=&-d_{ss}^c-3d_{sv}^c+\frac{d_{vs}^c}{2N_c}+\frac{3d_{vv}^c}{2N_c},  \nonumber \\ 
    d_{vv}&=&-d_{ss}^c+d_{sv}^c+\frac{d_{cs}^c}{2N_c}-\frac{d_{vv}^c}{2N_c}.
\end{eqnarray}
Both bases of $\mathcal{L}_{\psi\chi}$ will be employed in this study for convenience.
The Lagrangian in Eq.~(\ref{eqn:L4pt2}) is more convenient for matching with annihilation processes, while that in Eq.~(\ref{eqn:L4pt}) is advantageous for bound state calculations.

\section{Bi-linear operators}
\label{sec:twoq}

Any loop diagram in a perturbative quantum field theory
evaluated using dimensional regularization can be written as a function,
$F(\lbrace p\rbrace, \lbrace m\rbrace,\mu,\epsilon)$, where $\lbrace p\rbrace$
are the external four-momenta, $\lbrace m\rbrace$ are the external and internal
masses, $\mu$ is the 't~Hooft mass, and $d=4-2\epsilon$ is the space-time
dimension.

Let us then consider the radiative corrections to the
quark-gluon three-point vertex.
In QCD, this vertex can be expressed fully in terms of two form factors,
$F_1(q^2)$ and $F_2(q^2)$, defined by the irreducible three-point function,
\begin{equation}
    \Gamma_3^{\mathrm{QCD}}=-ig_sT^a\Bar{u}(p')\left[F_1(q^2)\gamma^{\mu} +iF_2(q^2)\frac{\sigma^{\mu\nu}q_{\nu}}{2m}\right]A_{\mu}^a(q)u(p)\,,
     \label{eqn:3ptvertQCD}
\end{equation}
where $p$ and $p'$ are the four-momenta of the incoming and outgoing quarks,
$q=p'-p$ is the four-momentum transfer from the gluon, and
$\sigma^{\mu\nu}=-\frac{i}{4}[\gamma^{\mu},\gamma^{\nu}]$.
We have just two form factors, as $\gamma_{\mu}$ and
$\sigma^{\mu\nu}q_{\nu}$ are the only Lorentz structures that appear in
QCD due to its nonchiral nature.
On the other hand, if one considers $\Gamma_3$ in the full SM, two additional
chiral Lorentz structures, with form factors $F_3(q^2)$ and $F_4(q^2)$, emerge,
\begin{equation}    \Gamma_3^{\mathrm{SM}}=\Gamma_3^{\mathrm{QCD}}-ig_sT^a\Bar{u}(p')\left[F_3(q^2)\gamma^{\mu}\gamma_{5}+F_4(q^2)\frac{q^{\mu}\gamma_5}{2m}\right]A_{\mu}^a(q)u(p)\,.
    \label{eqn:3ptvertSM}
\end{equation}
Moreover, the quark-photon three-point function will appear, albeit at sub-leading order, when considering the full SM with its associated form factors. As the methods for obtaining bi-linear nonrelativistic Lagrangian operators are analogous in both cases, we will focus our analysis on the leading quark-gluon vertex function and its associated bi-linear terms.

Evaluating the vertex and wave-function renormalization (WFR) Feynman diagrams in dimensional regularization, one finds that
the form factors $F_1(q^2)$ and $F_2(q^2)$ are UV and IR divergent
\cite{Manohar:1997qy}.
We can always expand our form factors, $F_i(q^2/m^2,\mu/m,\epsilon)$, as power
series in $q^2/m^2$ at fixed value of $\epsilon$ and then take the limit
$\epsilon\rightarrow 0$ to obtain an expression of the form
\begin{equation}
  F_{i}(q^2)=F_{i}\left[\frac{A_0}{\epsilon_{\mathrm{UV}}}+\frac{B_0}{\epsilon_{\mathrm{IR}}}+(A_0+B_0)\ln{\frac{\mu}{m}+D_0}\right]
+\frac{q^2}{m^2} F_i'\left[\frac{A_1}{\epsilon_{\mathrm{UV}}}+\frac{B_1}{\epsilon_{\mathrm{IR}}}+(A_1+B_1)\ln{\frac{\mu}{m}+D_1} \right]
+\mathcal{O}\left(\frac{q^4}{m^4}\right)\,,
\end{equation}
where we have introduced the short-hand notation
\begin{equation}
  F_i\equiv F_i(0)\,,\qquad
  F_i'\equiv \left.\frac{dF_i}{d(q^2/m^2)}\right|_{q^2=0}\,.
\label{eq:fexp}
\end{equation}
As usual, we label $\epsilon$ with the subscripts UV and IR to indicate whether
the divergence is ultraviolet or infrared, respectively.
UV divergences are canceled by renormalization counterterms, while IR
divergences cancel when a physical observable is considered.

The coefficients of the effective Lagrangian may be determined from the difference between the form factors in the full theory and the EFT.
More specifically, the nonanalytic terms in the form factors cancel in the
difference, while the analytic ones determine the Wilson coefficients of the
effective Lagrangian.
By inspection of the terms in the effective Lagrangian in Eq.~(\ref{eqn:L2pt}),
all of them contain at least one power of the gauge field, $A^{\mu}$.
Thus, the form factors at one loop are attainable by computing the three-point
scattering amplitudes with external quark lines on mass shell.

\subsection{Form factors}
\label{ssec:formfactors}

\begin{figure*}[t]
    \centering\captionsetup{justification=centering} 
\begin{center}
\fcolorbox{white}{white}{
\scalebox{0.45}{
    \begin{picture}(979,135) (33,-62)
    \SetWidth{3.0}
    \SetColor{Black}
    \Line[arrow,arrowpos=0.5,arrowlength=6.667,arrowwidth=2.667,arrowinset=0.2](34,-58)(124,-58)
    \Line[arrow,arrowpos=0.5,arrowlength=6.667,arrowwidth=2.667,arrowinset=0.2](124,-58)(241,-58)
    \Line[arrow,arrowpos=0.5,arrowlength=6.667,arrowwidth=2.667,arrowinset=0.2](241,-58)(332,-58)
    \PhotonArc[clock](180.5,-52.692)(72.694,-175.813,-364.187){7.5}{12.5}
    \Line[arrow,arrowpos=0.5,arrowlength=6.667,arrowwidth=2.667,arrowinset=0.2](374,-57)(461,-57)
    \Line[arrow,arrowpos=0.5,arrowlength=6.667,arrowwidth=2.667,arrowinset=0.2](461,-57)(579,-57)
    \Line[arrow,arrowpos=0.5,arrowlength=6.667,arrowwidth=2.667,arrowinset=0.2](579,-57)(671,-57)
    \Line[arrow,arrowpos=0.5,arrowlength=6.667,arrowwidth=2.667,arrowinset=0.2](711,-58)(801,-58)
    \Line[arrow,arrowpos=0.5,arrowlength=6.667,arrowwidth=2.667,arrowinset=0.2](801,-58)(919,-58)
    \Line[arrow,arrowpos=0.5,arrowlength=6.667,arrowwidth=2.667,arrowinset=0.2](919,-58)(1011,-58)
    \GluonArc[clock](860.69,-41.831)(75.842,-170.001,-371.537){7.5}{21}
    \Arc[dash,dashsize=2,clock](519.585,-44.595)(76.597,-170.68,-370.079)
    \Text(150,51)[lb]{\huge{\Black{$\gamma,W,Z$}}}
    \Text(500,52)[lb]{\huge{\Black{$H,\phi$}}}
  \end{picture}}
}
\end{center}
\caption{Feynman diagrams contributing to the quark self energy at one loop in
  the SM.} 
\label{fig:WFR}
\end{figure*}

At one loop in the SM, the vertex in Eq.~(\ref{eqn:3ptvertSM}) is endowed with
an on-shell WFR contribution as
\begin{equation}
    \hat{\Gamma}^{\rm{SM}}_3={\Gamma}^{\rm{SM}}_3-ig_sT^a\bar{u}(p')\gamma^{\mu}\left(\delta Z_V-\gamma_5\delta Z_A\right)A^a_{\mu}u(p)\,,
\end{equation}
where \cite{Bohm:1986rj}
\begin{eqnarray}
    \delta Z_V & =& -\Sigma_V(m^2)-2m^2\left[\Sigma_V'(m^2)+\Sigma_S'(m^2)\right]\,,\nonumber\\
    \delta Z_A &=& \Sigma_A(m^2)\,,
\label{eq:wfr1}
\end{eqnarray}
are defined in terms of the three scalar functions appearing in the self-energy
of the heavy quark,
\begin{equation}
\Sigma(p)=\slashed{p}\Sigma_V(p^2)+\slashed{p}\gamma_5\Sigma_A(p^2)+m\Sigma_S(p^2)\,,
\end{equation}
which arises from the Feynman diagrams shown in Fig.~\ref{fig:WFR}.

\begin{figure}[b]
\captionsetup{justification=centering}
\centering
\begin{center}
\scalebox{0.45}{
\fcolorbox{white}{white}{
 \begin{picture}(600,228) (65,-8)
    \SetWidth{1.0}
    \SetColor{Black}
    \Text(193,-13)[lb]{\Large{\Black{$(a)$}}}
    \Text(498,-11)[lb]{\Large{\Black{$(b)$}}}
    \SetWidth{2.0}
    \Line[arrow,arrowpos=0.5,arrowlength=6.667,arrowwidth=2.667,arrowinset=0.2](66,132)(131,132)
    \Line[arrow,arrowpos=0.5,arrowlength=6.667,arrowwidth=2.667,arrowinset=0.2](131,132)(192,132)
    \Line[arrow,arrowpos=0.5,arrowlength=6.667,arrowwidth=2.667,arrowinset=0.2](192,132)(257,132)
    \Line[arrow,arrowpos=0.5,arrowlength=6.667,arrowwidth=2.667,arrowinset=0.2](257,132)(319,132)
    \Gluon(194,130)(194,20){7.5}{9}
    \PhotonArc[clock](193.103,144.667)(65.342,-168.822,-372.073){7.5}{12.5}
    \Line[arrow,arrowpos=0.5,arrowlength=6.667,arrowwidth=2.667,arrowinset=0.2](368,181)(433,164)
    \Line[arrow,arrowpos=0.5,arrowlength=6.667,arrowwidth=2.667,arrowinset=0.2](563,164)(625,181)
    \Gluon(496,86)(561,163){7.5}{8}
    \Gluon(498,83)(499,20){7.5}{5}
    \Gluon(435,163)(497,85){7.5}{8}
    \Line[arrow,arrowpos=0.5,arrowlength=6.667,arrowwidth=2.667,arrowinset=0.2](433,164)(563,164)
  \end{picture}
}}
\end{center}
\caption{Feynman diagrams yielding (a) Abelian and (b) non-Abelian
 contributions to three-point matching coefficients in the SM. Wavy lines represent SM bosons.} 
\label{fig:Adiag}
\end{figure}

The total on-shell form factors at one-loop can be calculated from the Feynman diagrams depicted in Fig.~\ref{fig:Adiag}. We only present $F_1$, $F_2$, and $F_4$, as $F_3$ is related to $F_4$ via
\begin{equation}
      F_3(q^2)=F_3^{(0)}(q^2)-\delta Z_A  = -\frac{q^2}{4m^2}F_4(q^2),
\label{eq:f3}
\end{equation}
which is also enforced by current conservation. 

We present our results in the limit of large external on-shell top-quark mass, $m_t$, small internal bottom-quark mass from flavor changing, $m_b$, and small momentum squared, $q^2$, required for matching through $\mathcal{O}(1/m_t^3)$.
Specifically, we have
\begin{eqnarray}
  F_1(q^2)&=&1-\delta Z_V + F_1^{(a)}(q^2)+ F_1^{(b)}(q^2)
  \nonumber\\
  &=&1+\frac{\alpha_s}{\pi}\frac{q^2}{m_t^2}
  \left[-\frac{C_A}{8} \left(\frac{5}{3\varepsilon _{\mathrm{IR}}}
    +\frac{5}{6}\ln\frac{\mu^2}{m_t^2} +\frac{1}{2}\right)
      -C_F\left(\frac{1}{3 \varepsilon _{\mathrm{IR}}}
      +\frac{1}{6}\ln\frac{\mu^2}{m_t^2}+\frac{1}{8}\right)\right]
  \nonumber \\
  &&{}+\frac{\alpha}{\pi}\,\frac{q^2}{m_t^2}\left\{
  -\frac{2}{27}\left(\frac{2}{\epsilon_{\mathrm{IR}}}+\ln\frac{\mu^2}{m_Z^2}\right)
  -\frac{1}{144c_w^2}\left(\frac{17}{3}\ln\frac{m_Z^2}{m_t^2}+\frac{13}{2}\right)
\right.\nonumber \\
&&{}+\left.\frac{1}{192s_w^2}\left[4\left(\frac{m_t^2}{m_W^2}+3\right)
  \left(\ln\frac{m_b^2}{m_t^2}+i\pi\right)
  -4\ln\frac{m_Z^2}{m_t^2}
  -4\frac{m_t^2-3m_H^2}{m_W^2}\ln\frac{m_H^2}{m_t^2}
 -\frac{m_t^2-m_H^2}{m_W^2}+9\pi\frac{m_tm_H}{m_W^2}-10\right]\right\}\,,
  \nonumber\\
  F_2(q^2)&=&F_2^{(a)}(q^2)+ F_2^{(b)}(q^2)
  \nonumber\\
  &=&\frac{\alpha_s}{\pi}\left[
    \frac{C_A}{2}\left(\frac{1}{\epsilon_{\mathrm{IR}}}
    +\frac{1}{2}\ln\frac{\mu^2}{m_t^2}+1\right)
    +\frac{C_F}{2}\right]
  +\frac{\alpha_s}{\pi}\,\frac{q^2}{m_t^2}
  \left[\frac{C_A}{2}\left(\frac{1}{\epsilon_{\mathrm{IR}}}
    +\frac{1}{2}\ln\frac{\mu^2}{m_t^2}+\frac{1}{6}\right)
    +\frac{C_F}{12}\right]
     \nonumber\\
     &&{}+\frac{\alpha}{\pi}\left\{
     \frac{1}{144c_w^2}\left(9\ln\frac{m_Z^2}{m_t^2}+35\right)
     +\frac{1}{16 s_w^2}\left[
       \left(3\frac{m_H^2}{m_W^2}+1\right)\ln\frac{m_Z^2}{m_t^2}
       -6\frac{m_H^2}{m_W^2}\ln\frac{m_H^2}{m_t^2}
       +\frac{m_t^2}{m_W^2}-2\pi\frac{m_tm_H}{m_W^2}+7\right]
     \right\}
     \nonumber\\
     &&{}+\frac{\alpha}{\pi}\,\frac{q^2}{m_t^2}\left\{
     \frac{13}{432c_w^2}+\frac{1}{96s_w^2}\left[
       -2\frac{m_t^2}{m_W^2}\left(\ln\frac{m_b^2}{m_t^2}+i\pi\right)
       -6\frac{m_H^2}{m_W^2}\ln\frac{m_H^2}{m_t^2}
       +\frac{m_t^2}{m_W^2}-3\frac{m_H^2}{m_W^2}-3\pi\frac{m_tm_H}{m_W^2}+2\right]
     \right\}\,,
\nonumber \\
     F_4(q^2)&=&F_4^{(a)}(q^2)
     \nonumber\\
     &=&\frac{\alpha}{\pi}\left\{
     \frac{5}{144c_w^2}\left(6\ln\frac{m_Z^2}{m_t^2}+7\right)
     -\frac{1}{48s_w^2}\left[
    4\left(\frac{m_t^2}{m_W^2}-1\right)\left(\ln\frac{m_b^2}{m_t^2}+i\pi\right)
+6\ln\frac{m_Z^2}{m_t^2}+3\frac{m_t^2}{m_W^2}+27\right]\right\}
   \nonumber\\
   &&{}+\frac{\alpha}{\pi}\,\frac{q^2}{m_b^2}\,\frac{1}{60s_w^2}
   \left(\frac{m_t^2}{m_W^2}-1-\frac{m_W^2}{m_t^2}\right)
   +\frac{\alpha}{\pi}\,\frac{q^2}{m_t^2}\left\{
   \frac{1}{288c_w^2}\left(10\ln\frac{m_Z^2}{m_t^2}+21\right)
   \right.
   \nonumber\\
&&{}-\left.\frac{1}{480s_w^2}
   \left[2\left(7\frac{m_t^2}{m_W^2}+12\right)\left(\ln\frac{m_b^2}{m_t^2}+i\pi\right)+10\ln\frac{m_Z^2}{m_t^2}+19\frac{m_t^2}{m_W^2}+85\right]\right\}\,,
\label{eq:ff}
\end{eqnarray}
where we relate the Yukawa couplings to the EW coupling, $\alpha$, and the quark masses in the standard way~\cite{Denner:1991kt}, and we define the EW mixing angle as $c_w=\cos{\theta_w}=m_W/m_Z$ and $s_w=\sin{\theta_w}$.
The superscripts $(a)$ and $(b)$ of the form factors indicate the Abelian and non-Abelian contributions, respectively (see Fig.~\ref{fig:Adiag}).

Although the form factors in the limit presented above provide an adequate approximation for $m_t\gg m_W,m_Z,m_H \gg m_b,q^2$, another appropriate limit for the SM is $m_t\sim  m_W,m_Z,m_H \gg m_b,q^2$.
We leave a comparison of the various limits to future numerical studies.
The full expression, without any approximations, is included in an ancillary file posted along with our arXiv submission.

\subsection{Matching}
\label{ssec:matching}

To find the relationship between the full-theory form factors and the Wilson
coefficients for the scattering of a low-momentum heavy quark off a background
vector potential, we expand Eq.~(\ref{eqn:3ptvertSM}) in the nonrelativistic
limit and multiply by a factor of $\sqrt{m/E}$ for both the incoming and
outgoing quarks.

Adopting the momentum assignments from Eq.~(\ref{eqn:3ptvertQCD}), we are then
left with the effective interaction operator,
\begin{equation}
    -ig_sT^au_{\mathrm{NR}}^{\dagger}(\boldsymbol{p'})\left(A^{a}_0j^0-\boldsymbol{A}^a\cdot\boldsymbol{j}\right)u_{\mathrm{NR}}(\boldsymbol{p})\,,
    \label{eqn:effint}
\end{equation}
with current $j^\mu=(j^0,\boldsymbol{j})$.
This can thus be compared to the scattering amplitude of the effective
Lagrangian to relate the Wilson coefficients to the form factors.
We recalculate the nonrelativistic expansion of Eq.~(\ref{eqn:effint}) in QCD
and confirm the well-established result
\cite{Manohar:1997qy,Kinoshita:1995mt}, {\it i.e.}\ we find the
time component of the current to be
\begin{equation}
    j^0=F_1(q^2)\left[ 1-\frac{1}{8m^2}\boldsymbol{q}^2+\frac{i}{4m^2}\boldsymbol{\sigma}\cdot(\boldsymbol{p'}\times\boldsymbol{p})\right] + F_2(q^2)\left[-\frac{1}{4m^2}\boldsymbol{q}^2+\frac{1}{2m^2}\boldsymbol{\sigma}\cdot(\boldsymbol{p'}\times\boldsymbol{p})\right]\,,
\end{equation}
and its spatial component to be
\begin{eqnarray}
  \boldsymbol{j}&= & F_1(q^2)\left[ \frac{1}{2m}(\boldsymbol{p}+\boldsymbol{p'})+\frac{i}{2m}\boldsymbol{\sigma}\times\boldsymbol{q}-\frac{i}{8m^3}(\boldsymbol{p}^2+\boldsymbol{p'}^2)\boldsymbol{\sigma}\times\boldsymbol{q}
    - \frac{1}{16m^3}(\boldsymbol{p'}^2-\boldsymbol{p}^2)\boldsymbol{q}+\frac{i}{16m^3}(\boldsymbol{p}^2-\boldsymbol{p'}^2)\boldsymbol{\sigma}\times(\boldsymbol{p}+\boldsymbol{p'})
    \right. \nonumber \\ &&{} \left.
    -\frac{1}{8m^3}(\boldsymbol{p'}^2+\boldsymbol{p}^2)(\boldsymbol{p'}+\boldsymbol{p})\right]
  + F_2(q^2)\left[ \frac{i}{2m}\boldsymbol{\sigma}\times\boldsymbol{q}-\frac{i}{16m^3}\boldsymbol{q}^2\boldsymbol{\sigma}\times\boldsymbol{q}-\frac{1}{16m^3}\boldsymbol{q}^2(\boldsymbol{p}+\boldsymbol{p'})
    -\frac{1}{16m^3}(\boldsymbol{p'}^2-\boldsymbol{p}^2)\boldsymbol{q}
    \right.\nonumber \\ &&{} \left.
     -\frac{i}{8m^3}(\boldsymbol{p'}^2-\boldsymbol{p}^2)\boldsymbol{\sigma}\times(\boldsymbol{p'}+\boldsymbol{p})
    +\frac{i}{8m^3}\boldsymbol{\sigma}\cdot(\boldsymbol{p'}+\boldsymbol{p})(\boldsymbol{p'}\times\boldsymbol{p})\right]\,.
\end{eqnarray}
This can then be compared with the relevant subset of the Hamiltonian in
Eq.~(\ref{eqn:L2pt}),
\begin{eqnarray}
  \mathcal{H}_{\psi,\chi}&\supset&\psi^{\dagger}\left[ g_sA^0-c_2\frac{g_s}{2m}\boldsymbol{A}\cdot(\boldsymbol{p}'+\boldsymbol{p})-c_D\frac{g_s}{8m^2}\boldsymbol{q}^2A^0-ic_F\frac{g_s}{2m}\boldsymbol{A}\cdot(\boldsymbol{\sigma}\times\boldsymbol{q})
-c_D\frac{g_s}{16m^3}(\boldsymbol{p}'^2-\boldsymbol{p}^2)\boldsymbol{q}\cdot\boldsymbol{A}    
    \right.  \nonumber \\  &&{}
     + ic_S \frac{g_s}{4m^2}\boldsymbol{\sigma}\cdot(\boldsymbol{p}'\times\boldsymbol{p})A^0
    + ic_S \frac{g_s}{16m^3}(\boldsymbol{p}'^2-\boldsymbol{p}^2)\boldsymbol{A}\cdot\boldsymbol{\sigma}\times(\boldsymbol{p}'+\boldsymbol{p}) 
    +i(c_{W_1}-c_{W_2})\frac{g_s}{8m^3}(\boldsymbol{p}'^2+\boldsymbol{p}^2)\boldsymbol{A}\cdot(\boldsymbol{\sigma}\times\boldsymbol{q})
    \nonumber \\  &&{}
    +ic_{W_2}\frac{g_s}{8m^3}\boldsymbol{q}^2\boldsymbol{A}\cdot(\boldsymbol{\sigma}\times\boldsymbol{q}) -c_M\frac{g_s}{8m^3}(\boldsymbol{p}'^2-\boldsymbol{p}^2)\boldsymbol{A}\cdot\boldsymbol{q}
    -ic_q \frac{g_s}{8m^3}\boldsymbol{\sigma}\cdot(\boldsymbol{p}'+\boldsymbol{p})\boldsymbol{A}\cdot(\boldsymbol{p}'\times\boldsymbol{p})
    \nonumber \\  &&{}
    + \left. c_M\frac{g_s}{8m^3}\boldsymbol{q}^2\boldsymbol{A}\cdot(\boldsymbol{p}'+\boldsymbol{p})\right]\psi + (\mathrm{c.c.}, \psi\leftrightarrow\chi)
  \nonumber\\ 
        &\equiv& g_s\psi^{\dagger}\left( A^0j^0-\boldsymbol{A}\cdot\boldsymbol{j} \right) \psi + (\mathrm{c.c.},\psi\leftrightarrow\chi)\,.
\end{eqnarray}
Matching the Lorentz structures provides one with the following relations
between the Wilson coefficients and the form factors:
\begin{eqnarray}
    c_0&=&c_2=c_4=F_1\,,\nonumber\\
    c_F &=&F_1+F_2\,,\nonumber\\
    c_D&=&F_1+2F_2+8F_1'\,,\nonumber\\
    c_S&=&F_1+2F_2\,,\nonumber\\
    c_{W_1}&=&F_1+\frac{1}{2}F_2+4F_1'+4F_2'\,,\nonumber\\
    c_{W_2}&=&\frac{1}{2}F_2+4F_1'+4F_2'\,,\nonumber\\
    c_q&=&F_2\,,\nonumber\\
    c_M&=&\frac{1}{2}F_2+4F_1'\,.
    \label{eqn:wilform}
\end{eqnarray}
Moreover, re-parametrization invariance imposes constraints on the Wilson coefficients \cite{Luke:1992cs},
\begin{eqnarray}
    c_0&=&c_2=c_4=F_1\,,\nonumber\\
    c_S &=&2c_F-1\,,\nonumber\\
    c_{W_2}&=&c_{W_1}-1\,,\nonumber\\
    c_q&=&c_F-1\,,\nonumber\\
    2c_M&=&c_D-c_F\,,
\end{eqnarray}
which are satisfied by Eq.~(\ref{eqn:wilform}) and also reflected by the
numbers in Table~\ref{tab:i} to be discussed in Section~\ref{sec:discussion}. 

The relations between the form factors and Wilson coefficients remain
unchanged upon allowing further interactions from the SM.
To see this, we take the small-$q^2$ limit of Eq.~(\ref{eqn:3ptvertSM}),
after substituting Eq.~(\ref{eq:f3}), by expanding the form factors in $q^2$.
Retaining terms through $\mathcal{O}(q^2/m^3)$, we thus have
\begin{eqnarray}
  \Gamma_3^{\rm{SM}}&=& -ig_s\bar{u}(p')\left[\left(F_1+\frac{q^2}{m^2}F_1'\right)\gamma_{\mu}+\frac{i}{2m}\left(F_2+\frac{q^2}{m^2}F_2'\right)\sigma^{\mu\nu}q_{\nu}+\frac{1}{2m}F_4\left(q^{\mu}-\frac{q^2}{2m}\gamma^{\mu}\right)\gamma_5
    +\frac{q^2}{2m^3}F_4'q^{\mu}\gamma_5\right]
  \nonumber\\
  &&{}\times A_{\mu}(q)u(p)\,,
   \label{eqn:nlovert}
\end{eqnarray}
where we have used the notation of Eq.~(\ref{eq:fexp}).
Upon expansion, a low energy EFT can be deduced by comparison with the
expanded vertex in Eq.~(\ref{eqn:nlovert}). At LO in $q^2$, the EFT has the form
\begin{equation}
    \mathcal{L}_{\rm{rel}}^{(0)}=\bar{Q}\left( i\slashed{D}F_1-m_{\rm{ren}}+ig_sF_2\frac{\sigma^{\mu\nu}}{2m}G_{\mu\nu}-\frac{g_s}{4m^2}F_4D_{\mu}G^{\mu\nu}\gamma_{\nu}\gamma^5\right) Q\,,
    \label{eqn:relLagr0}
\end{equation}
with field strength $G^{\mu\nu}\equiv \frac{i}{g_s}[D^{\mu},D^{\nu}]$, $F_1=1$, and $F_2,F_4=\mathcal{O}(\alpha_s,\alpha)$. Moreover, using the gluonic equations of motion, which dictate conservation of color charge,
\begin{equation}
    [D_{\mu},G^{\mu\nu}]=g_s\sum_i \bar{q}_i\gamma^{\nu}T^aq_i T^a\,,
    \label{eqn:coleom}
\end{equation}
the last term in Eq.~(\ref{eqn:relLagr0}) can be re-written as a four-quark operator,
\begin{equation}
    \mathcal{L}_{\rm{rel}}^{(0)}\supset -\frac{g_s^2}{4m^2}F_4\sum_i\bar{q}_i\gamma_{\nu}T^a{q}_i\bar{Q}\gamma^{\nu}\gamma_5T^a{Q}\,.
\label{eq:ewop}
\end{equation}
Similarly, at $\mathcal{O}(q^2)$, one can deduce the EFT Lagrangian by comparison with the expanded vertex and so obtains
\begin{eqnarray}
  \mathcal{L}_{\rm{rel}}^{(1)}&=&\bar{Q}\left\lbrace \frac{1}{m^2}g_sD_{\mu}G^{\mu\nu}\gamma_{\nu}F_1'+ig_sF_2'\frac{\sigma^{\mu\nu}}{2m^3}\left( iD_{\rho}\left[iD^{\rho},G_{\mu\nu}\right]+\left[G_{\mu\nu},iD^{\rho}\right]iD_{\rho}\right)
  \right.\nonumber\\{}&&-\left.
  \frac{g_s}{4m^4}F_4'\left( iD_{\mu}\left[iD^{\mu},D_{\mu}G^{\mu\nu}\gamma_{\nu}\gamma^5\right]
  +\left[D_{\mu}G^{\mu\nu}\gamma_{\nu}\gamma^5,iD^{\mu}\right]iD_{\mu}\right)\right
    \rbrace Q\,.
    \label{eqn:relLagr1}
\end{eqnarray}
Again using Eq.~(\ref{eqn:coleom}), one may re-express the last operator in Eq.~(\ref{eqn:relLagr1}) as a four-quark operator, thus eliminating it from the current. One can next employ the HQET expansion of the bi-spinors to obtain the
nonrelativistic Lagrangian. 
Starting from the relativistic Lagrangian,
\begin{equation}
   \mathcal{L}_{\rm{rel}}=\bar{Q}\left( i\slashed{D}-m+iX\right)Q\,,
   \label{eqn:Xlag}
\end{equation}
where $X$ are virtual corrections from the full theory at
$\mathcal{O}(\alpha_s,\alpha)$, decomposing the four-component spinor field
$Q=e^{-imv\cdot x}(h_v+H_v)$ into the large upper and small lower components $h_v=e^{imv\cdot x}P_ +Q$ and $H_v=e^{imv\cdot x}P_ -Q$, where $P_{\pm}=(1\pm \slashed{v})/2$, so that
 $\slashed{v}h_v=h_v$ and $\slashed{v}H_v=-H_v$, and defining
$D_{\perp}^{\mu}=D^{\mu}-v^{\mu}v\cdot D$, we obtain
\begin{equation}
  \mathcal{L}=
\bar{h}_vi(v\cdot D+X)h_v-\bar{H}_v[i(v\cdot D-X)+2m]H_v+\bar{h}_vi(\slashed{D}_{\perp}+X)H_v+\bar{H}_vi(\slashed{D}_{\perp}+X)h_v\,.
\label{eq:deco}
\end{equation}
Notice that Eq.~(\ref{eq:deco}) implies a decomposition of $X$,
\begin{equation}
X={X}_{++}+{X}_{+-}+{X}_{-+}+{X}_{--}\,,
\end{equation}
where ${X}_{ab}=P_a{X}P_b$ with $a,b=+,-$.
Plugging the equation of motion for $\bar{H}_v$,
\begin{equation}
    H_v=\frac{1}{i(v\cdot D-X_{--})+2m}i(\slashed{D}_{\perp}+X_{-+})h_v\,,
\end{equation}
into Eq.~(\ref{eq:deco}), we find
\begin{align}
    \mathcal{L}={}&\bar{h}_vi(v\cdot D+X_{++})h_v+\bar{h}_vi(\slashed{D}_{\perp}+X_{+-})\frac{1}{i(v\cdot D-X_{--})+2m}i(\slashed{D}_{\perp}+X_{-+})h_v \nonumber\\
    ={}&\bar{h}_vi(v\cdot D+X_{++})h_v+\frac{1}{2m}\sum_{n=0}^{\infty}\bar{h}_vi(\slashed{D}_{\perp}+X_{+-})\left(\frac{i(-v\cdot D+X_{--})}{2m}\right)^ni(\slashed{D}_{\perp}+X_{-+})h_v\,.
\label{eq:hvhv}
\end{align}

We focus on the $F_{4}$ contribution to $X$, whose structure is unfamiliar from
pure QCD.
It comes as the $\mathcal{O}(\alpha,1/m^2)$ term
$iX^{\rm{ch}}\equiv-g_s F_4D_{\mu}G^{\mu\nu}\gamma_{\nu}\gamma^5/(4m^2)$ in Eq.~(\ref{eqn:relLagr0}), where the
subscript ``ch'' is to indicate its chiral nature, involving $\gamma_5$.
Inserting this term in Eq.~(\ref{eq:hvhv}) and expanding through
$\mathcal{O}(1/m^3)$, one obtains the covariant chiral extension,
\begin{align}
    \mathcal{L}_{\rm{ch}}={}&\bar{h}_v X^{\rm{ch}}_{++} h_v+\frac{1}{2m}\bar{h}_v\left( i\slashed{D}_{\perp}X^{\rm{ch}}_{-+}+iX^{\rm{ch}}_{+-}\slashed{D}_{\perp}\right) h_v+\mathcal{O}\left(\frac{1}{m^4}\right) \nonumber\\ 
    ={}& \bar{h}_v\left(-\frac{F_4}{m^2}g_s D_{\mu}G^{\mu\nu}+\frac{F_4}{8m^3}g_s\lbrace iD_{\perp,\rho},D_{\mu}G^{\mu\nu}v_{\nu}\rbrace\right)\gamma^{\rho}_{\perp}\gamma_5 h_v\,,
\end{align}
with
\begin{align}
iX_{++}^{\rm{ch}}=iX_{--}^{\rm{ch}}={}& -\frac{F_4}{4m^2}g_sD_{\mu}G^{\mu\nu}\gamma_{\perp,\nu}\gamma_5\,, \nonumber\\ 
iX_{+-}^{\rm{ch}}=-iX_{-+}^{\rm{ch}}={}& -\frac{F_4}{4m^2}g_sD_{\mu}G^{\mu\nu}v_{\nu}\gamma_5\,, 
\end{align}
where $\gamma_{\perp,\mu}=\gamma_{\mu}-v_{\mu}\slashed{v}$ and the chirality is made explicit by the appearance of the factor $\gamma_5$ in each term. Note that the contributions proportional to $F'_4$ in Eq.~(\ref{eqn:relLagr1}) lead to $\mathcal{O}(\alpha,1/m^4)$ terms. In the reference frame where $v^\mu=(1,\boldsymbol{0})$, 
\begin{equation}
\mathcal{L}^{\rm{ch}}_{\psi,\chi}={\psi}^\dagger\left(-\frac{F_4}{m^2}g_s D_{\mu}G^{\mu i}+\frac{F_4}{8m^3}g_s\lbrace i{D}^i,{D}_{j}G^{j0}\rbrace\right){\sigma}_i \psi + (\mathrm{c.c.}, \psi\leftrightarrow\chi)  +\mathcal{O}\left(\frac{1}{m^4},\frac{g_s^2}{m^3}\right)\,.
\label{eqn:hqetch}
\end{equation}
Thus, to completely account for leading EW corrections, the bi-linear Lagrangian $\mathcal{L}_{\psi,\chi}$ in Eq.~(\ref{eqn:L2pt}) must be modified as
\begin{equation}
\mathcal{L}_{\psi,\chi}\rightarrow \mathcal{L}_{\psi,\chi}+\mathcal{L}^{\rm{ch}}_{\psi,\chi},
\end{equation}
and the Wilson coefficients of $\mathcal{L}_{\psi,\chi}$ have to be endowed with
their EW corrections, which enter through the form factors $F_1$ and $F_2$. The additional operator in Eq.~\eqref{eqn:hqetch} leads to additional EFT vertices. Using the notation of Ref.~\cite{Pineda:2011dg}, we illustrate the single gluon emission rules in Fig.~\ref{fig:ChFrules}.

\begin{figure*}[tb]
\centering\captionsetup{justification=centering} 
\begin{center}
\scalebox{0.45}{
\fcolorbox{white}{white}{
  \begin{picture}(648,444) (45,-11)
    \SetWidth{2.0}
    \SetColor{Black}
    \Line[arrow,arrowpos=0.5,arrowlength=6.667,arrowwidth=2.667,arrowinset=0.2](49,395)(145,350)
    \Line[arrow,arrowpos=0.5,arrowlength=6.667,arrowwidth=2.667,arrowinset=0.2](145,350)(243,394)
    \SetWidth{1.0}
    \Vertex(145,350){6.708}
    \SetWidth{2.0}
    \Line[dash,dashsize=2,arrow,arrowpos=0.5,arrowlength=6.667,arrowwidth=2.667,arrowinset=0.2](146,349)(146,241)
    \Text(48,411)[lb]{\Large{\Black{$\beta$}}}
    \Text(243,410)[lb]{\Large{\Black{$\alpha$}}}
    \Text(48,375)[lb]{\Large{\Black{$\boldsymbol{p}$}}}
    \Text(244,373)[lb]{\Large{\Black{$\boldsymbol{p'}$}}}
    \Text(167,269)[lb]{\Large{\Black{$\boldsymbol{q}$}}}
    \Text(241,306)[lb]{\Large{\Black{$-ig_s\frac{F_4}{m^2}q^0\boldsymbol{q}\cdot\boldsymbol{\sigma}T^a_{\alpha\beta}$}}}
    \Text(146,227)[lb]{\Large{\Black{$a,0$}}}
    \Line[arrow,arrowpos=0.5,arrowlength=6.667,arrowwidth=2.667,arrowinset=0.2](47,152)(143,107)
    \Line[arrow,arrowpos=0.5,arrowlength=6.667,arrowwidth=2.667,arrowinset=0.2](143,107)(242,152)
    \Line[dash,dashsize=2,arrow,arrowpos=0.5,arrowlength=6.667,arrowwidth=2.667,arrowinset=0.2](144,106)(144,-2)
    \Text(46,168)[lb]{\Large{\Black{$\beta$}}}
    \Text(241,172)[lb]{\Large{\Black{$\alpha$}}}
    \Text(46,132)[lb]{\Large{\Black{$\boldsymbol{p}$}}}
    \Text(241,129)[lb]{\Large{\Black{$\boldsymbol{p'}$}}}
    \Text(165,26)[lb]{\Large{\Black{$\boldsymbol{q}$}}}
    \Text(239,63)[lb]{\Large{\Black{$-ig_s\frac{F_4}{8m^3}\left(\boldsymbol{p'}+\boldsymbol{p}\right)\cdot\boldsymbol{\sigma}\boldsymbol{q}^2T^a_{\alpha\beta}$}}}
    \Text(144,-16)[lb]{\Large{\Black{$a,0$}}}
    \Line[arrow,arrowpos=0.5,arrowlength=6.667,arrowwidth=2.667,arrowinset=0.2](464,153)(560,108)
    \Line[arrow,arrowpos=0.5,arrowlength=6.667,arrowwidth=2.667,arrowinset=0.2](560,108)(655,152)
    \Text(463,169)[lb]{\Large{\Black{$\beta$}}}
    \Text(656,170)[lb]{\Large{\Black{$\alpha$}}}
    \Text(463,133)[lb]{\Large{\Black{$\boldsymbol{p}$}}}
    \Text(656,134)[lb]{\Large{\Black{$\boldsymbol{p'}$}}}
    \Text(582,27)[lb]{\Large{\Black{$\boldsymbol{q}$}}}
    \Text(656,64)[lb]{\Large{\Black{$-ig_s\frac{F_4}{8m^3}\left(\boldsymbol{p'}+\boldsymbol{p}\right)\cdot\boldsymbol{\sigma}q^0\boldsymbol{q}^jT^a_{\alpha\beta}$}}}
    \Text(561,-15)[lb]{\Large{\Black{$a,j$}}}
    \Line[arrow,arrowpos=0.5,arrowlength=6.667,arrowwidth=2.667,arrowinset=0.2](466,395)(562,350)
    \Line[arrow,arrowpos=0.5,arrowlength=6.667,arrowwidth=2.667,arrowinset=0.2](562,350)(658,396)
    \SetWidth{1.0}
    \Vertex(562,350){6.708}
    \Text(465,411)[lb]{\Large{\Black{$\beta$}}}
    \Text(657,412)[lb]{\Large{\Black{$\alpha$}}}
    \Text(465,375)[lb]{\Large{\Black{$\boldsymbol{p}$}}}
    \Text(656,379)[lb]{\Large{\Black{$\boldsymbol{p'}$}}}
    \Text(584,269)[lb]{\Large{\Black{$\boldsymbol{q}$}}}
    \Text(658,306)[lb]{\Large{\Black{$-ig_s\frac{F_4}{m^2}\left(\boldsymbol{q}^2\delta_{ij}-\boldsymbol{q}_i\boldsymbol{q}_j\right)\boldsymbol{\sigma}^iT^a_{\alpha\beta}$}}}
    \Text(563,227)[lb]{\Large{\Black{$a,j$}}}
    \SetWidth{2.0}
    \Photon(563,349)(565,248){7.5}{5}
    \Photon(561,107)(563,1){7.5}{5}
    \SetWidth{1.0}
    \GBox(135,101)(153,116){0.882}
    \GBox(556,102)(574,117){0.882}
  \end{picture}}
}
\end{center}
\caption{Additional NRQCD vertices from EW corrections in Feynman gauge with $q=p'-p$. The dotted and wavy lines correspond to transverse and longitudinal gluons. In Coulomb gauge the bottom-right vertex vanishes.
}
\label{fig:ChFrules}
\end{figure*}

Notice that we have written the chiral-symmetric HQET Lagrangian in the special frame, with
$v^\mu=(1,\boldsymbol{0})$, and employed the notation of
Ref.~\cite{Kinoshita:1995mt}.
However, one can also rewrite Eq.~(\ref{eqn:L2pt}) in an arbitrary frame as
\begin{eqnarray}
  \mathcal{L}_v&=&\Bar{h}_v\left\lbrace c_0iD\cdot v -c_2\frac{D_{\perp}^2}{2m}+c_4\frac{D_{\perp}^4}{8m^3}-g_sc_F\frac{\sigma_{\mu\nu}G^{\mu\nu}}{4m}
  -g_sc_D\frac{v^{\mu}[D_{\perp}^{\nu}G_{\mu\nu}]}{8m^2} +ig_sc_S\frac{v_{\lambda}\sigma_{\mu\nu}\lbrace D_{\perp}^{\mu},G^{\nu\lambda} \rbrace}{8m^2}
  \right.\nonumber \\ &&{}
  +g_sc_{W_1}\frac{\lbrace D_{\perp}^{2},\sigma_{\mu\nu}G^{\mu\nu} \rbrace}{16m^3} -g_sc_{W_2}\frac{ D_{\perp}^{\lambda}\sigma_{\mu\nu}G^{\mu\nu} D_{\perp\lambda}}{8m^3}
  +g_sc_q\frac{\sigma^{\mu\nu}(D^{\lambda}_{\perp}G_{\lambda\mu}D_{\perp\nu}+D_{\perp\nu}G_{\lambda\mu}D_{\perp}^{\lambda}-D^{\lambda}_{\perp}G_{\mu\nu}D_{\perp\lambda})}{8m^3} \nonumber \\ &&{}- \left. ig_sc_M\frac{D_{\perp\mu}[D_{\perp\nu}G^{\mu\nu}]+[D_{\perp\nu}G^{\mu\nu}]D_{\perp\mu}}{8m^3} \right\rbrace h_v\,.
    \label{eqn:L2ptI}
\end{eqnarray}

Alternatively, one can then proceed as in Eq.~(\ref{eqn:effint}) with the nonrelativistic expansion and deduce the nonrelativistic Lagrangian along with Wilson coefficients in terms of $F_4$ and $F_4'$ up to $\mathcal{O}(1/m^3)$. The current is shifted as $j \to j +j'$,
where $j'$ includes the new form factor, $F_4$, and its extended Lorentz structures.
For the time component of the current extension, one so obtains
\begin{eqnarray}
    j'^0& =& 
    F_4(q^2)\left[-\frac{\boldsymbol{q}^2}{8m^3}\boldsymbol{\sigma}\cdot (\boldsymbol{p'}+\boldsymbol{p})-\frac{1}{8m^3}\boldsymbol{\sigma}\cdot \boldsymbol{q}(\boldsymbol{p'}^2-\boldsymbol{p}^2)\right]\,,
\end{eqnarray}
and for its spatial component
\begin{eqnarray}
    \boldsymbol{j'}&=& 
    F_4(q^2)\left( \frac{\boldsymbol{q}^2}{4m^2}\boldsymbol{\sigma} -\frac{1}{4m^2}\boldsymbol{q} \boldsymbol{\sigma}\cdot\boldsymbol{q}\right)\,.
\end{eqnarray}

\section{Four-quark operators}
\label{sec:fourq}

To achieve the four-quark matching, we follow a procedure originally outlined
for the QCD case in Ref.~\cite{Pineda:1998kj}, reproduce the results obtained
there, and extend them to the EW case.
In fact, owing to the absence of derivative terms in the four-quark portion of
our effective Lagrangian, in Eqs.~(\ref{eqn:L4pt}) and (\ref{eqn:L4pt2}), we
are entitled to consider the dimensionally regulated $S$ matrix elements with
the four external heavy quarks being exactly on mass shell and at rest.

Thus, in the leading nonrelativistic limit considered, the external four-component spinors can be written as
\begin{equation}
    u(p)=\sqrt{2m}\left( {\begin{array}{c}
\psi  \\
0
\end{array} } \right),\quad v(-p)=\sqrt{2m}\left( {\begin{array}{c}
0  \\
\chi
\end{array} } \right)\,,
\end{equation}
with Pauli spinors $\psi$ ($\chi$) representing a heavy quark (antiquark) and the usual normalization as given in Ref.~\cite{Beneke:2013jia}. Thus, unlike the bi-linear EW corrections up to $\mathcal{O}(1/m^3)$, new explicit operators induced by chirality are not yet apparent in the four-quark matching at $\mathcal{O}(1/m^2)$.

We exclusively use the $\overline{\mathrm{MS}}$ scheme to regularize the
appearing singularities, which are of UV and IR types.
At first sight, one would also expect Coulomb singularites to emerge. 
In fact, $S$ matrix elements of such heavy-heavy systems are known to exhibit a
unique IR behavior, which gives rise to Coulomb poles and the standard
nonrelativistic weak-coupling bound states.
However, this is only true if the Coulomb singularity is regularized by the
relative momentum of the heavy quarks or by assigning an infinitesimal mass to
the exchange gluon.
However, this odd powerlike IR divergence does not surface in dimensional
regularization.
The EFT exhibits an identical IR behavior, which is consistently quenched by
dimensional regularization.

We also use the $\overline{\mathrm{MS}}$ scheme to renormalize the basic
parameters.
However, we use the on-shell scheme to define the WFR constants, as in
Eq.~(\ref{eq:wfr1}), to comply with the Lehmann-Symanzik-Zimmermann
\cite{Lehmann:1954rq,Lehmann:1957zz} condition for asymptotic states in the
first place, without having to apply finite adjustments.

\subsection{Equal-mass case}
\label{ssec:equalm}

In the equal-mass case of QCD, The Feynman diagrams which contribute to the matching and are not already taken into account by the bi-linear Lagrangian in Eq.~(\ref{eqn:L2pt}) are the QCD box diagrams of Fig.~\ref{fig:scatt} and Fig.~\ref{fig:annih}. Thus, there are both scattering contributions given by $d_{xy}$ with $m_1=m_2=m$ and annihilation contributions to $d_{xy}^c$, already starting at tree level.
We recover the annihilation matching coefficients in pure QCD obtained in Ref.~\cite{Pineda:1998kj},
\begin{eqnarray}
  d_{ss}^c&=&\alpha_s^2C_F(-C_A+2C_F)\left(\ln2-1-i\frac{\pi}{2}\right)\,,
  \nonumber\\  
    d_{sv}^c&=&0\,, \nonumber\\
    d_{vs}^c&=&\alpha_s^2\left(\frac{3}{2}C_A-4C_F\right)
\left(\ln2-1-i\frac{\pi}{2}\right)\,,
    \nonumber\\
    d_{vv}^c&=&-\pi\alpha_s+\alpha_s^2\left[
      \frac{1}{6}\left(-\frac{11}{2}C_A+n_f+1\right)
      \ln\frac{\mu^2}{m^2}
      -\frac{109}{36}C_A+4C_F-\frac{n_f}{3}
\left(\ln2-\frac{5}{6}-i\frac{\pi}{2}\right)      
+\frac{4}{9}\right]\,,
\label{eq:emqcd}
\end{eqnarray}
and our results agree with  previous calculations~\cite{Bodwin:1994jh,Pineda:1998kj}.
The $\mathcal{O}(\alpha)$ and $\mathcal{O}(\alpha\alpha_s)$ EW corrections to these coefficients, denoted as $\Delta d_{xy}^c$, are presented below in the limit $m_t  \gg m_W,m_Z,m_H \gg m_b$, for the sake of compactness. We have through $\mathcal{O}({M_{{\rm EW}}/m_t,m_b/M_{{\rm EW}}})$,
\begin{eqnarray}
  \Delta d_{ss}^c&=&-\pi\alpha\frac{m_t^2}{4s_w^2m_W^2}
  -\alpha\alpha_s\frac{C_F}{2s_w^2}\left[\frac{3}{8}\left(2\frac{m_t^2}{m_W^2}
    +\frac{1}{2c_w^2}\right)
    \ln\frac{\mu^2}{m_t^2}-\frac{m_t^2}{2m_W^2}
    +\frac{1}{4c_w^2}\right]\,,
\nonumber\\ 
\Delta d_{sv}^c&=&-\pi\alpha\frac{1}{16}
\left(\frac{25}{9c_w^2}+\frac{1}{s_w^2}\right)
+\alpha\alpha_s\frac{C_F}{4}\left(\frac{25}{9c_w^2}+\frac{1}{s_w^2}\right)\,,
  \nonumber\\
  \Delta d_{vs}^c&=&\alpha\alpha_s\frac{1}{4}
  \left(\frac{25}{9c_w^2}+\frac{1}{s_w^2}\right)
  \left(1+i\frac{\pi}{2}-\ln2\right)\,,
  \nonumber\\
  \Delta d_{vv}^c&=&-\alpha\alpha_s
  \left\{\frac{\ln2}{3s_w^2}\left[\frac{1}{m_W^2}\left(m_t^2-\frac{m_H^2}{4}\right)-1\right]-\frac{m_H^2}{4m_W^2s_w^2}\ln\frac{m_H^2}{m_t^2} 
    -\frac{1}{4}\left[\frac{25}{9c_w^2}+\frac{5}{3s_w^2}+\frac{7m_t^2}{3m_W^2s_w^2}\right] 
   \right.\nonumber\\&&{}\left. 
   +\pi\left[\frac{1}{s_w^2}\left(-\frac{7 m_W}{192 c_w m_t}+\frac{5
   m_t}{24 c_w m_W}+\frac{157
   m_H^3}{768 m_t m_W^2}+\frac{i
   m_H^2}{24 m_W^2}+\frac{m_t^3}{2
   m_H m_W^2}-\frac{11 m_H m_t}{48
   m_W^2}-\frac{i m_t^2}{3 m_W^2}
   \right)\right.\right.\nonumber\\&&{}\left.\left. +\frac{113 m_Z^3}{1728 m_t m_W^2}+\frac{2
   m_t m_Z}{9 m_W^2}-\frac{11 m_Z}{27
   m_t}-\frac{8 m_t}{9 m_Z}\right]\right\}\,.
    \label{eq:emewf}
\end{eqnarray}
The imaginary parts in $\Delta d_{vs}^c$ and $\Delta d_{vv}^c$ displayed above are related to the tree level cross-sections $t\bar t\to g\gamma+g Z$ and $t\bar t \to g H + g Z$ respectively by the optical theorem. We have checked that the expected relations hold.

Additionally, in the SM, scattering amplitudes contribute to the matching coefficients $d_{xy}$ as shown in Fig.~\ref{fig:scatt}, and through  is given by,
\begin{eqnarray}
\Delta d_{ss}&=& \pi\alpha\frac{m_t^2}{m_W^2}\left[
\frac{1}{s_w^2}\left(\frac{m_t^2}{m_H^2}+\frac{1}{4}\right)
  -\frac{16}{9}c_w^2+\frac{4}{9}\right]-\alpha\alpha_sC_F\frac{m_t^2}{m_W^2}\frac{m_t^2}{m_H^2s_w^2}\left\{
1-\frac{3}{2}
\ln\frac{\mu^2}{m_t^2}\right\}\,,
\nonumber\\
\Delta d_{sv}&=&-\pi\alpha\frac{m_t^2}{4s_w^2m_W^2}
+\alpha\alpha _sC_F\frac{m_t^2}{4s_w^2m_W^2}\,,
\nonumber\\
\Delta d_{vs}&=&-\alpha\alpha _s\left\{
\frac{8}{9}\left(\frac{1}{\epsilon_{\mathrm{IR}}}+\ln\frac{\mu^2}{m_t^2}\right)
-\frac{16}{27}-\frac{5}{24s_w^2}+\frac{19}{216c_w^2}+\frac{m_H^2}{10m_W^2s_w^2}-\frac{m_t^2}{m_W^2s_w^2}\left(\frac{5}{6}-\frac{1}{2}\ln\frac{m_H^2}{m_t^2}\right)  \right.\nonumber\\
  &&{}\left.
+\pi\left[\frac{1}{s_w^2}\left(\frac{2 m_t^5}{m_H^3 m_W^2 }+\frac{3m_t^3}{4 m_H m_W^2 }-\frac{9 m_Hm_t}{64 m_W^2 }+\frac{m_t^3}{2 m_W^2m_Z }+\frac{5 m_tm_Z}{16 m_W^2}-\frac{5m_H^3}{512m_tm_W^2}-\frac{21m_Z}{256m_t}\right)
\right.\right.\nonumber\\
  &&{}\left.\left.
+\frac{8 m_t^3}{9 m_W^2m_Z}
-\frac{32 m_t^3}{9 m_Z^3}
+\frac{m_t m_Z}{9m_W^2}-\frac{4 m_t}{9 m_Z}-\frac{5m_Z}{4m_t}-\frac{101m_Z^3}{256m_tm_W^2}\right]+\frac{1}{9}\left(\frac{9}{8s_w^2}+\frac{25}{8c_w^2}-8\right)\ln\frac{m_Z^2}{m_t^2}\right\}\,,
\nonumber\\ 
\Delta d_{vv}&=&-\alpha\alpha _s\left\{
\frac{1}{72}\left(\frac{35}{c_w^2}+\frac{19}{s_w^2}\right)+\frac{5m_H^2}{36m_W^2s_w^2}+\frac{m_t^2}{2m_W^2s_w^2}-\frac{m_H^2}{12m_W^2s_w^2}\ln\frac{m_H^2}{m_t^2}+\frac{1}{24c_w^2s_w^2}\ln\frac{m_Z^2}{m_t^2}  
\right.\nonumber\\
  &&{}\left.
+\pi\left[\frac{1}{s_w^2}\left(\frac{3 m_H^3}{128 m_t
   m_W^2}-\frac{m_t^3}{3 m_H
   m_W^2}-\frac{m_H m_t}{8
   m_W^2}-\frac{32 m_t m_W^2}{27
   m_Z^3}-\frac{8 m_t m_Z}{27
   m_W^2}-\frac{9 m_Z}{128 m_t}+\frac{40
   m_t}{27 m_Z}\right)
      \right.\right.\nonumber\\
  &&{}\left.\left.
   \frac{1}{s_w^4}\left(\frac{5 c_w m_t}{24 m_W}-\frac{5
   m_t}{24 c_w m_W}\right)-\frac{9m_Z^3}{128m_W^2m_t}+\frac{32m_t}{27m_Z}\right]\right\}\,.
   \label{eq:emew}
\end{eqnarray}
The IR poles appearing in $\Delta d_{vs}$ should cancel against UV poles in the non-relativistic effective theory calculation in physical amplitudes. We have checked that this is actually so. We have also checked that the remaining $\mu$-dependence (unrelated to IR poles) in $\Delta d_{ss} $ and $\Delta d_{ss}^c$ corresponds to the running of the top Yukawa coupling and fixes it to the scale $m_t$, much in the same way as the $\mu$-dependence in (\ref{eq:emqcd}) corresponds to the running of $\alpha_s$ and fixes it to the scale $m_t$ \cite{Pineda:1998kj}.
The full result through $\mathcal{O}(\alpha^2)$ is included as an ancillary file in our arXiv submission.

Notice that imaginary parts frequently appearing in Wilson coefficients are related by the optical theorem to inelastic cross sections, which are unattainable from nonrelativistic theory alone.
In particular, the partial widths of the decays of heavy-quarkonium states into light hadrons are also implicated in such imaginary parts.

\subsection{Unequal-mass case}
\label{ssec:unequalm}

\begin{figure*}[tb]
\centering\captionsetup{justification=centering} 
\begin{center}
\scalebox{0.4}{
\fcolorbox{white}{white}{
 \begin{picture}(871,331) (43,-25)
    \SetWidth{3.0}
    \SetColor{Black}
    \Line[dash,dashsize=2,arrow,arrowpos=0.5,arrowlength=6.667,arrowwidth=2.667,arrowinset=0.2](801,296)(801,168)
    \Line[arrow,arrowpos=0.5,arrowlength=6.667,arrowwidth=2.667,arrowinset=0.2](673,296)(801,296)
    \Line[arrow,arrowpos=0.5,arrowlength=6.667,arrowwidth=2.667,arrowinset=0.2](801,296)(913,296)
    \Line[arrow,arrowpos=0.5,arrowlength=6.667,arrowwidth=2.667,arrowinset=0.2](913,168)(785,168)
    \Line[arrow,arrowpos=0.5,arrowlength=6.667,arrowwidth=2.667,arrowinset=0.2](785,168)(673,168)
    \GOval(801,232)(20,21)(0){0.882}
    \Line[arrow,arrowpos=0.5,arrowlength=6.667,arrowwidth=2.667,arrowinset=0.2](49,296)(161,296)
    \Line[arrow,arrowpos=0.5,arrowlength=6.667,arrowwidth=2.667,arrowinset=0.2](161,296)(273,296)
    \Line[arrow,arrowpos=0.5,arrowlength=6.667,arrowwidth=2.667,arrowinset=0.2](273,168)(161,168)
    \Line[arrow,arrowpos=0.5,arrowlength=6.667,arrowwidth=2.667,arrowinset=0.2](161,168)(49,168)
    \Line[dash,dashsize=2](161,296)(161,168)
    \Photon(735,107)(847,-21){7.5}{9}
    \Line[arrow,arrowpos=0.5,arrowlength=6.667,arrowwidth=2.667,arrowinset=0.2](671,107)(735,107)
    \Line[arrow,arrowpos=0.5,arrowlength=6.667,arrowwidth=2.667,arrowinset=0.2](735,107)(847,107)
    \Line[arrow,arrowpos=0.5,arrowlength=6.667,arrowwidth=2.667,arrowinset=0.2](847,107)(911,107)
    \Line[arrow,arrowpos=0.5,arrowlength=6.667,arrowwidth=2.667,arrowinset=0.2](911,-21)(847,-21)
    \Line[arrow,arrowpos=0.5,arrowlength=6.667,arrowwidth=2.667,arrowinset=0.2](847,-21)(735,-21)
    \Line[arrow,arrowpos=0.5,arrowlength=6.667,arrowwidth=2.667,arrowinset=0.2](735,-21)(671,-21)
    \Photon(847,107)(735,-21){7.5}{9}
    \Line[arrow,arrowpos=0.5,arrowlength=6.667,arrowwidth=2.667,arrowinset=0.2](367,107)(431,107)
    \Line[arrow,arrowpos=0.5,arrowlength=6.667,arrowwidth=2.667,arrowinset=0.2](431,107)(543,107)
    \Line[arrow,arrowpos=0.5,arrowlength=6.667,arrowwidth=2.667,arrowinset=0.2](543,107)(591,107)
    \Line[arrow,arrowpos=0.5,arrowlength=6.667,arrowwidth=2.667,arrowinset=0.2](591,-21)(543,-21)
    \Line[arrow,arrowpos=0.5,arrowlength=6.667,arrowwidth=2.667,arrowinset=0.2](543,-21)(431,-21)
    \Line[arrow,arrowpos=0.5,arrowlength=6.667,arrowwidth=2.667,arrowinset=0.2](431,-21)(367,-21)
    \Photon(431,107)(431,-21){7.5}{6}
    \Photon(527,107)(527,-21){7.5}{6}
    \Photon(433,296)(529,296){7.5}{5}
    \Line[arrow,arrowpos=0.5,arrowlength=6.667,arrowwidth=2.667,arrowinset=0.2](364,296)(433,296)
    \Line[arrow,arrowpos=0.5,arrowlength=6.667,arrowwidth=2.667,arrowinset=0.2](433,296)(481,232)
    \Line[arrow,arrowpos=0.5,arrowlength=6.667,arrowwidth=2.667,arrowinset=0.2](481,232)(529,296)
    \Line[arrow,arrowpos=0.5,arrowlength=6.667,arrowwidth=2.667,arrowinset=0.2](529,296)(593,296)
    \Line[arrow,arrowpos=0.5,arrowlength=6.667,arrowwidth=2.667,arrowinset=0.2](593,168)(481,168)
    \Line[arrow,arrowpos=0.5,arrowlength=6.667,arrowwidth=2.667,arrowinset=0.2](481,168)(369,168)
    \Line[dash,dashsize=2](481,232)(481,168)
    \Line[arrow,arrowpos=0.5,arrowlength=6.667,arrowwidth=2.667,arrowinset=0.2](44,107)(113,107)
    \Line[arrow,arrowpos=0.5,arrowlength=6.667,arrowwidth=2.667,arrowinset=0.2](209,107)(273,107)
    \Line[arrow,arrowpos=0.5,arrowlength=6.667,arrowwidth=2.667,arrowinset=0.2](273,-21)(161,-21)
    \Line[arrow,arrowpos=0.5,arrowlength=6.667,arrowwidth=2.667,arrowinset=0.2](161,-21)(49,-21)
    \Line[dash,dashsize=2](161,43)(161,-21)
    \Line[arrow,arrowpos=0.5,arrowlength=6.667,arrowwidth=2.667,arrowinset=0.2](115,107)(208,107)
    \Line[dash,dashsize=2](208,106)(163,43)
    \Line[dash,dashsize=2](114,106)(162,41)
  \end{picture}}
}
\end{center}
\caption{Scattering diagrams relevant for the matching to the nonrelativistic four-quark operators at $\mathcal{O}(1/m^2)$ up to one-loop order in the SM. Wavy lines represent bosons and dashed lines exclusively massive bosons. The incoming and outgoing quarks are on mass shell and exactly at rest. 
}
\label{fig:scatt}
\end{figure*}

As for the unequal quark mass case in QCD, annihilation diagrams do not contribute. Thus, the Feynman diagrams which contribute to the matching and are not already taken into account by the bi-linear Lagrangian in Eq.~(\ref{eqn:L2pt}) are the QCD box diagrams of Fig.~\ref{fig:scatt}.
We recalculate the matching coefficients appearing in Eq.~(\ref{eqn:L4pt}) at one loop in QCD and confirm the results of Ref.~\cite{Pineda:1998kj},
\begin{eqnarray}
  d_{ss}&=&C_F\left(\frac{C_A}{2}-C_F\right)\alpha_s^2\left[
      \frac{1}{\epsilon_{\mathrm{IR}}}+\ln\frac{\mu^2}{m_1m_2}-\frac{1}{3}
+\frac{m_1^2+m_2^2}{2\left(m_1^2-m_2^2\right)}\ln\frac{m_1^2}{m_2^2}\right]\,,
  \nonumber\\ 
  d_{sv}&=&C_F\left(\frac{C_A}{2}-C_F\right)\alpha_s^2\frac{m_1m_2}{m_1^2-m_2^2}
  \ln\frac{m_1^2}{m_2^2}\,,
  \nonumber\\ 
  d_{vs}&=&\alpha_s^2\left\{\left(-\frac{3}{4}C_A+2C_F\right)\left[
      \frac{1}{\epsilon_{\mathrm{IR}}}+\ln\frac{\mu^2}{m_1m_2}-\frac{1}{3}
      +\frac{m_1^2+m_2^2}{2\left(m_1^2-m_2^2\right)}\ln\frac{m_1^2}{m_2^2}\right]
  \right.
  \nonumber\\
  &&{}-\left.\frac{C_A}{4m_1m_2}\left[\left(m_1^2+m_2^2\right)
    \left(\frac{1}{\epsilon_{\mathrm{IR}}}+\ln\frac{\mu^2}{m_1m_2}-\frac{10}{3}\right)
    +\frac{m_1^4+m_2^4}{2\left(m_1^2-m_2^2\right)}\ln\frac{m_1^2}{m_2^2}\right]
  \right\}\,,
\nonumber\\
d_{vv}&=&\alpha_s^2\left\{-\frac{C_A}{4}\left[
  \frac{1}{\epsilon_{\mathrm{IR}}}+\ln\frac{\mu^2}{m_1m_2}-3
    +\frac{m_1^2+m_2^2}{2\left(m_1^2-m_2^2\right)}\ln\frac{m_1^2}{m_2^2}\right]
+\left(-\frac{3}{4}C_A+2C_F\right)
\frac{m_1m_2}{m_1^2-m_2^2}\ln\frac{m_1^2}{m_2^2}\right\}\,.
\label{eq:umqcd}
\end{eqnarray}

\begin{figure*}[tb]
\centering\captionsetup{justification=centering} 
\begin{center}
\scalebox{0.4}{
\fcolorbox{white}{white}{
 \begin{picture}(932,332) (46,-30)
    \SetWidth{3.0}
    \SetColor{Black}
    \Photon(768,108)(848,-20){7.5}{8}
    \Line[arrow,arrowpos=0.5,arrowlength=6.667,arrowwidth=2.667,arrowinset=0.2](688,108)(768,108)
    \Line[arrow,arrowpos=0.5,arrowlength=6.667,arrowwidth=2.667,arrowinset=0.2](768,108)(768,-20)
    \Line[arrow,arrowpos=0.5,arrowlength=6.667,arrowwidth=2.667,arrowinset=0.2](768,-20)(688,-20)
    \Line[arrow,arrowpos=0.5,arrowlength=6.667,arrowwidth=2.667,arrowinset=0.2](848,108)(928,108)
    \Line[arrow,arrowpos=0.5,arrowlength=6.667,arrowwidth=2.667,arrowinset=0.2](928,-20)(848,-20)
    \Line[arrow,arrowpos=0.5,arrowlength=6.667,arrowwidth=2.667,arrowinset=0.2](848,-20)(848,108)
    \Photon(768,-20)(848,108){7.5}{8}
    \Line[arrow,arrowpos=0.5,arrowlength=6.667,arrowwidth=2.667,arrowinset=0.2](656,300)(720,236)
    \Line[arrow,arrowpos=0.5,arrowlength=6.667,arrowwidth=2.667,arrowinset=0.2](720,236)(656,172)
    \Photon(720,236)(800,236){7.5}{4}
    \Photon(832,236)(912,236){7.5}{4}
    \Line[arrow,arrowpos=0.5,arrowlength=6.667,arrowwidth=2.667,arrowinset=0.2](912,236)(976,300)
    \Line[arrow,arrowpos=0.5,arrowlength=6.667,arrowwidth=2.667,arrowinset=0.2](976,172)(912,236)
    \GOval(814,233)(22,22)(0){0.882}
    \Photon(448,108)(544,108){7.5}{5}
    \Photon(448,-20)(544,-20){7.5}{5}
    \Line[arrow,arrowpos=0.5,arrowlength=6.667,arrowwidth=2.667,arrowinset=0.2](384,108)(448,108)
    \Line[arrow,arrowpos=0.5,arrowlength=6.667,arrowwidth=2.667,arrowinset=0.2](448,108)(448,-20)
    \Line[arrow,arrowpos=0.5,arrowlength=6.667,arrowwidth=2.667,arrowinset=0.2](448,-20)(384,-20)
    \Line[arrow,arrowpos=0.5,arrowlength=6.667,arrowwidth=2.667,arrowinset=0.2](544,108)(608,108)
    \Line[arrow,arrowpos=0.5,arrowlength=6.667,arrowwidth=2.667,arrowinset=0.2](608,-20)(544,-20)
    \Line[arrow,arrowpos=0.5,arrowlength=6.667,arrowwidth=2.667,arrowinset=0.2](544,-20)(544,108)
    \Line[arrow,arrowpos=0.5,arrowlength=6.667,arrowwidth=2.667,arrowinset=0.2](64,108)(128,44)
    \Line[arrow,arrowpos=0.5,arrowlength=6.667,arrowwidth=2.667,arrowinset=0.2](128,44)(64,-20)
    \Photon(128,44)(208,44){7.5}{4}
    \Photon(208,44)(240,12){7.5}{3}
    \Photon(208,44)(240,76){7.5}{3}
    \Line[arrow,arrowpos=0.5,arrowlength=6.667,arrowwidth=2.667,arrowinset=0.2](240,76)(272,108)
    \Line[arrow,arrowpos=0.5,arrowlength=6.667,arrowwidth=2.667,arrowinset=0.2](240,12)(240,76)
    \Line[arrow,arrowpos=0.5,arrowlength=6.667,arrowwidth=2.667,arrowinset=0.2](272,-20)(240,12)
    \Line[arrow,arrowpos=0.5,arrowlength=6.667,arrowwidth=2.667,arrowinset=0.2](48,300)(112,236)
    \Line[arrow,arrowpos=0.5,arrowlength=6.667,arrowwidth=2.667,arrowinset=0.2](112,236)(48,172)
    \Photon(112,236)(224,236){7.5}{6}
    \Line[arrow,arrowpos=0.5,arrowlength=6.667,arrowwidth=2.667,arrowinset=0.2](224,236)(288,300)
    \Line[arrow,arrowpos=0.5,arrowlength=6.667,arrowwidth=2.667,arrowinset=0.2](288,172)(224,236)
    \Line[arrow,arrowpos=0.5,arrowlength=6.667,arrowwidth=2.667,arrowinset=0.2](352,300)(416,236)
    \Line[arrow,arrowpos=0.5,arrowlength=6.667,arrowwidth=2.667,arrowinset=0.2](416,236)(352,172)
    \Photon(416,236)(528,236){7.5}{6}
    \Line[arrow,arrowpos=0.5,arrowlength=6.667,arrowwidth=2.667,arrowinset=0.2](528,236)(560,268)
    \Line[arrow,arrowpos=0.5,arrowlength=6.667,arrowwidth=2.667,arrowinset=0.2](560,268)(592,300)
    \Line[arrow,arrowpos=0.5,arrowlength=6.667,arrowwidth=2.667,arrowinset=0.2](592,172)(560,204)
    \Line[arrow,arrowpos=0.5,arrowlength=6.667,arrowwidth=2.667,arrowinset=0.2](560,204)(528,236)
    \Photon(567,276)(567,197){6}{5}
  \end{picture}}
}
\end{center}
\caption{Annihilation diagrams relevant for the matching to the nonrelativistic four-quark operators at $\mathcal{O}(1/m^2)$  up to one-loop order. Wavy lines represent SM bosons. The incoming and outgoing quarks are on mass shell and exactly at rest.}
\label{fig:annih}
\end{figure*}

In the full SM, the quarks can no longer be generic, and we thus calculate the leading EW corrections specifically for the heaviest quarks, top and bottom. The matching coefficients are now determined by all Feynman diagrams in Figs.~\ref{fig:scatt} and \ref{fig:annih} since annihilation through $W$-exchange can occur. However, due to the fact that $m_b\ll M_{\rm EW}$, the binding energy of the system is much less than the top quark width, and thus, a bound state will not form before top decay. In light of this, we restrict ourselves to presenting solely the tree-level contribution for illustration purposes. At LO the unexpanded coefficients are given by,
\begin{eqnarray}
\Delta d_{ss}&=& \pi\alpha\frac{m_tm_b}{m_W^2}\left\{
\frac{1}{s_w^2}\left(\frac{m_tm_b}{m_H^2}-\frac{1}{4}\right)
  +\frac{8}{9}c_w^2+\frac{1}{9}\right\}\,,
\nonumber\\
\Delta d_{sv}&=&\pi\alpha\frac{m_tm_b}{4s_w^2m_W^2}\,,
\nonumber\\
\Delta d_{vs}&=&d_{vv}=0.
\label{eq:umew}
\end{eqnarray}
Moreover, in the SM, annihilation amplitudes are permitted through $W$ boson exchange and contribute to the matching coefficients, $d_{xy}^c$, as shown in Fig.~\ref{fig:annih}, and are given by,
\begin{eqnarray}
  \Delta d_{ss}^c&=&-\pi\alpha\frac{m_t m_b}{2s_w^2m_W^2}\,,
\nonumber\\ 
\Delta d_{sv}^c&=&-\pi  \alpha\frac{  m_t m_b}{2s_w^2((m_b+m_t)^2-m_W^2)}\,,
  \nonumber\\
  \Delta d_{vs}^c&=&d_{vv}^c=0.
\label{eq:umewf}
\end{eqnarray}

\section{Discussion}
\label{sec:discussion}

As in the original studies within pure QCD, our final results for the matching coefficients still contain IR divergences, extracted as poles in $\epsilon_{\mathrm{IR}}$ using dimensional regularization.
These IR divergences will be canceled by similar corrections due to the real radiation of soft gluons and photons when specific physical observables are considered.
Of course, the real radiative corrections need to be evaluated using dimensional regularization as well.
To be able to separately study the numerical sizes of the universal and process dependent corrections, it is useful to perform an $\overline{\mathrm{MS}}$ subtraction of the IR divergences, which boils down to dropping the $1/\epsilon_{\mathrm{IR}}$ terms in our expressions.
In the following, it is understood that this manipulation has been carried out.

The EW expansion is different from the QCD one---the expansion parameter being $\alpha$, rather than $\alpha_s$---and
its IR safety is indeed guaranteed by itself. The IR divergences of EW origin present in the various coefficient functions are due to virtual photons. They will be cancelled by similar corrections due to real soft-photon radiation once particular processes are calculated using the Feynman rules derived from the extended NRQCD Lagrangian. This is
analogous to the pure QCD case considered so far, where real soft-gluon corrections to specific processes cancel the IR singularities in Eqs.~(\ref{eq:ff}), (\ref{eq:umqcd}), and (\ref{eq:emqcd}). While the IR cancellations proceed independently in the QCD and EW sectors at the lowest nontrivial orders, they become intertwined at higher orders when gluons and photons can become soft simultaneously. Specifically, the IR divergences displayed in Eqs.~(\ref{eq:umew}) and (\ref{eq:emew}) are mixed QCD and EW effects, proportional to $\alpha\alpha_s$. The purely weak contributions, involving $W$ and $Z$ boson exchanges only, given in Eqs.~(\ref{eq:umewf}) and (\ref{eq:emewf}), respectively, are IR
safe as expected. Notice also that the matching coefficient of the only parity violating operator we find at this order, given in Eq.~(\ref{eq:ewop}), is also IR safe; see $F_4(0)$ in Eq.~(\ref{eq:ff}).

We are now in a position to explore the numerical significance of our results.
We consider the full set of EW corrections to the matching coefficients of the
two- and four-quark operators in Eqs.~(\ref{eqn:L2pt}) and (\ref{eqn:L4pt}),
focusing on their real parts, and compare with the well-known QCD results;
our results are presented in Tables~\ref{tab:i} and \ref{tab:ii},
respectively.
To maximize the accuracy, we avoid taking any limits, but use the full
expressions through the considered order.
We choose the renormalization scale to be $\mu=m_Z$, put $m_1=m_t(m_Z)$,
$m_2=m_b(m_Z)$, and $\alpha_s=\alpha_s(m_Z)$, and adopt the SM parameter values
from the latest Review of Particle Physics~\cite{ParticleDataGroup:2022pth}.
We recall that

\begin{table*}[t]
\captionsetup{justification=centering}
\begin{tabular}{|c||c|c|c|c|c|c|c|c|c|c|}
\hline
Coeff.\ & $c_{0,2,4}$& $c_F$& $c_D$& $c_S$& $c_{{{W}}_1}$& $c_{{{W}}_2}$& $c_q$& $c_M$ & $F_4$\\
\hline
QCD & 1 & 1.0447 & 1.1935 & 1.0893 & 0.9809 & $-0.01909$ & 0.0447 & 0.0744 & 0\\
\hline 
EW & 0 & 0.0007& $-0.0870$ & 0.0015 & $-0.0172$ & $-0.0172$ & 0.0007 & $-0.0439$ & $0.0143$ \\
\hline
\end{tabular}
\caption{\label{tab:i}Matching coefficients of the bi-linear quark operators in Eq.~(\ref{eqn:L2pt}) for the top quark, evaluated at renormalization scale $\mu=m_Z$ using SM parameter values from Ref.~\cite{ParticleDataGroup:2022pth}.}
\end{table*}

\begin{table*}[t]
\captionsetup{justification=centering}
\begin{tabular}{|c||c|c|c|c|c|c|c|c|}
\hline
Coeff.\ ($m_1=m_2$) & $d_{ss}$ & $d_{sv}$ & $d_{vs}$ & $d_{vv}$ & $d_{ss}^c$ & $d_{sv}^c$ & $d_{vs}^c$ & $d_{vv}^c$\\
\hline
QCD (LO+NLO) & 0.0019 & 0.0031 & 0.0833 & 0.0617 & $-0.0019$ & 0 & $-0.0036$ & $-0.3592$\\
\hline 
EW (LO) & 1.0180 & $-0.1290$ & 0 & 0 & $-0.1290$ & $-0.0121$ & 0 & 0\\
\hline
EW (NLO) & $-0.0704$ & $-0.0168$ & $-0.3794$ & $0.0622$ & $0.0099$ & $0.0004$ & $0.0006$ & $-0.0289$\\
\hline
EW (LO+NLO) & $0.9476$ & $-0.1458$ & $-0.3794$ & $0.0622$ & $-0.1191$ & $-0.0117$ & $0.0006$ & $-0.0289$\\
\hline
\hline
Coeff.\ ($m_1\neq m_2$)& $d_{ss}$ & $d_{sv}$ & $d_{vs}$ & $d_{vv}$ & $d_{ss}^c$ & $d_{sv}^c$ & $d_{vs}^c$ & $d_{vv}^c$\\
\hline
QCD (LO+NLO) & 0.0171 & 0.0006 & $-0.8997$ & -0.0076 & 0 & 0 & 0 & 0\\
\hline 
EW (LO) & $-0.0004$ & 0.0036 & 0 & 0 & $-0.0073$ & $-0.0018$ & 0 & 0\\
\hline
\end{tabular}
\caption{\label{tab:ii}Matching coefficients of the four-quark operators in Eq.~(\ref{eqn:L4pt}), for $t\bar{t}\rightarrow t\bar{t}$ ($m_1=m_2$) and $t\bar{b}\rightarrow t\bar{b}$ ($m_1\ne m_2$), evaluated at renormalization scale $\mu=m_Z$ using SM parameter values from Ref.~\cite{ParticleDataGroup:2022pth}.}
\end{table*}

We begin by considering the Wilson coefficients $c_i$ and $F_4$ of the two-quark operators, in Table~\ref{tab:i}.
Inspecting Table~\ref{tab:i}, we observe that the EW corrections alter the
Wilson coefficients significantly at the chosen renormalization scale.
Moreover, these corrections widely vary in size depending on the coefficient under consideration, and this provides further credence to the lack of reliability of naive order-of-magnitude estimates.
As for the new parity violating operators, they come equipped with nonnegligible matching coefficients, of order of magnitude similar to the ones of the parity preserving operators.
On the other hand, the matching coefficients of the four-quark operators vary even more strongly in both the QCD and EW sectors, as is evident from Table~\ref{tab:ii}. This is exemplified in the equal-mass case of $t\bar{t}\rightarrow t\bar{t}$, where EW corrections dominate in a majority of the matching coefficients. This is mainly due to the appearance of tree-level EW contributions. For instance, in the case of $d_{ss}$, the dominant EW contribution arises from tree-level Higgs boson exchange. We note, however, that these coefficients are heavily suppressed in the Lagrangian by the $1/m_t^2$ prefactor. Thus, the bi-linear QCD/QED terms dominate at the level of the EFT when considering $2\rightarrow 2$ processes. 

The EW corrections to top-quark pair production in $e^+e^-$ annihilation at threshold have been considered at NNLO in Ref.~\cite{Beneke:2017rdn}. Our results not only provide necessary ingredients for the full NNNLO calculation, but also show that the size of some of those ingredients is comparable to the size of the NNLO corrections, and in some cases even larger, for instance in the coefficients $d_{ss}$ and $d_{vs}$.
Notice also that, if one counts $\alpha\sim \alpha_s^2$, our results for the Wilson coefficients  match the precision of the QCD two-loop calculation of Ref.~\cite{Gerlach:2019kfo}.

If one instead considers $b\bar{b}\rightarrow b\bar{b}$, the difference between QCD and EW corrections becomes far more pronounced due to the lack of explicit numerator factors of $m_t$ in the full-theory amplitudes.
The effect of $m_t$ is already apparent in the unequal-mass case of $t\bar{b}\rightarrow t\bar{b}$, where the matching coefficients are more than one order of magnitude suppressed versus the $t\bar{t}\rightarrow t\bar{t}$ case. However, in QCD, annihilation diagrams are forbidden in this case, and thus $d^c_{xy}$ obtains only EW contributions through $W$ boson exchange.
Thus, in the nonrelativistic regime, our results further justify the necessity of EW corrections in precision calculations, in particular when considering processes involving the top quark.

We end by noting that our results have been obtained with the help of the programming language \texttt{Mathematica} accompanied by the package \texttt{FeynCalc}~\cite{Shtabovenko:2016sxi} to compute the necessary amplitudes and to deal with the algebra.
Furthermore, we employed subpackages of \texttt{FeynCalc}, such as \texttt{FeynHelpers}~\cite{Shtabovenko:2016whf}, which reduces the amplitudes and provides explicit expressions for one-loop scalar integrals by connecting the reduction package \texttt{fire}~\cite{Smirnov:2019qkx} with the analytic scalar-integral program \texttt{Package-X}~\cite{Patel:2016fam}.
Lastly, we employed the \texttt{FeynOnium} subpackage, which comes equipped with functions for dealing with calculations in the nonrelativistic limit~\cite{Brambilla:2020fla}.

\section{Conclusion}
\label{sec:conclusion}

In this paper, the matching coefficients of the NRQCD Lagrangian were computed at one loop, through order $\mathcal{O}(1/m^3)$ for the bi-linear operators and through order $\mathcal{O}(1/m^2)$ for the four-quark operators, within QCD as the full theory, confirming previous results.
The Lagrangian was then extended to include the leading QCD plus EW and purely EW corrections at one loop, of which various limits were presented and discussed.
Extending the NRQCD Lagrangian into the EW, just the photon appears as an additional field propagating relativistically, besides the gluon and the light quarks.
All the other EW fields, including the $H$, $W$, and $Z$ bosons, only show up in the coefficient functions. By the same token, the covariant derivatives receive an additional term proportional to the photon field. The additional Feynman rules thus generated are shown in Fig.~\ref{fig:ChFrules}. The velocity counting rules for the operators go unchanged, but the relative sizes of $v$, $\alpha_s(m_t)$, $\alpha$, and other EW parameters must be properly accommodated in the counting; see, e.g., Ref.~\cite{Beneke:2017rdn}.

A crucial result of our paper is that the terms of the original NRQCD Langrangian are not enough to capture all the new EW features, because certain symmetries that are manifest in QCD, such as parity and charge conugation, are broken in the SM. This makes it necessary to extend the original NRQCD Lagrangian by structures that are not amenable from within
pure QCD. Specifically, new parity violating operators were found to be necessary for the two-quark terms in the effective Lagrangian. The new terms arose due to the SM being parity violating, and new Lorentz structures emerged that are not present in the nonrelativistic limit of QCD. Thus, the matching coefficients accompanying the parity violating terms exhibited EW corrections in pure form. In the study of the four-quark operators, both the cases of equal and unequal external-heavy-quark masses were considered.

We rounded off by comparing all the matching coefficients for a particular renormalization scale with and without EW corrections, and found the contributions from the EW regime to be relevant.
Therefore, we recommend that these contributions be included in future high-precision studies that employ heavy-quark effective theories.

\vspace{1cm}

{\bf Acknowledgments:}
We thank A.~V.~Manohar for his helpful comments, and O.~L.~Veretin and Z.-G.~He for their technical advice.
This work of B.A.K. was supported in part by the German Research Foundation DFG through Grant No.\ KN~365/12-1.
The work of J.S. was supported in part by the Generalitat de Catalunya through
Grant No.\ 2021-SGR-249 and by the Ministerio de Ciencia, Innovaci\'on y Universidades through Projects No.\ PID2019-105614GB-C21, PID2019-110165GB-I00, and
CEX2019-000918-M.


\end{document}